\documentclass[preprint,12pt]{elsarticle}

\usepackage{graphicx}
\graphicspath{{./}} 
\usepackage{float} 
\usepackage{subfigure}
\usepackage{amsmath} 
\usepackage{amssymb}
\usepackage{amsthm}
\usepackage{amsfonts,bm}
\usepackage{mathptmx}
\usepackage{geometry}
\usepackage{titlesec}

\usepackage{booktabs}

\usepackage{caption}
\usepackage{hyperref}
\hypersetup{hypertex=true, colorlinks=true, linkcolor=blue, anchorcolor=blue, citecolor=blue}

\usepackage[normalem]{ulem}

\journal{Renewable Energy}

\usepackage{color}

\begin{document}

\begin{frontmatter}



\title{\LARGE\bfseries A three-dimensional dynamic mode decomposition analysis of wind farm flow aerodynamics}


\author[I1]{Xuan Dai}
\author[I1]{Da Xu}
\author[I1]{Mengqi Zhang}
\author[I2]{Richard J.A.M. Stevens}
\address[I1]{Department of Mechanical Engineering, National University of Singapore, 10 Kent Ridge Crescent, 119260, Singapore}
\address[I2]{Physics of Fluids Group, Max Planck Center Twente for Complex Fluid Dynamics, J. M. Burgers Center for Fluid Dynamics, and MESA Research Institute, University of Twente, P.O.Box 217, 7500 AE Enschede, The Netherlands}

\begin{abstract}
High-fidelity large-eddy simulations are suitable to obtain insight into the complex flow dynamics in extended wind farms. In order to better understand these flow dynamics, we use dynamic mode decomposition (DMD) to analyze and reconstruct the flow field in large-scale numerically simulated wind farms by large-eddy simulations (LES). Different wind farm layouts are considered, and we find that a combination of horizontal and vertical staggering leads to improved wind farm performance compared to traditional horizontal staggering. We analyze the wind farm flows using the amplitude selection (AP) and sparsity-promoting (SP method) DMD approach. We find that the AP method tends to select modes with a small length scale and a high frequency, while the SP method selects large coherent structures with low frequency. The latter are somewhat reminiscent of modes obtained using proper orthogonal decomposition (POD). We find that a relatively limited number of SP-DMD modes is sufficient to accurately reconstruct the flow field in the entire wind farm, whereas the AP-DMD method requires more modes to achieve an accurate reconstruction. Thus, the SP-DMD method has a smaller performance loss compared to the AP-DMD method in terms of the reconstruction of the flow field.
\end{abstract}




\end{frontmatter}



\section{Introduction}
\label{S:1}
Renewable and sustainable energy plays an increasingly important role in modern society's development \cite{RN391,RN393,RN390}. Wind energy is one of the fastest-growing forms of renewable energy and will form a central part of the future energy system. It draws extensive attention from academia and industry, not only because of its fast technological advancement but also for business opportunities \cite{RN395}. Most wind energy is produced in large wind farms in which wind turbines are clustered together. Unfortunately, this leads to wake effects, which can affect the performance of large offshore wind farms by 10\% to 20\% \cite{RN396}. Therefore, it is essential to understand these wake interactions and to further develop wind farm control strategies that can be employed to mitigate wake effects in extended wind farms \cite{stevens2017flow}.

This work will use high-fidelity large-eddy simulations to study the flow dynamics in extended wind farms. Such simulation cannot be used directly to design wind farm controllers as they are too computationally intensive. However, these simulations capture the relevant flow dynamics accurately, which can be studied using popular flow decomposition methods \cite{Hussain1983,Berkooz1993} such as proper orthogonal decomposition (POD) and dynamic mode decomposition (DMD). While POD enables one to find the most energetic time-averaged flow structures \cite{Berkooz1993} DMD allows one to capture the flow structures at different frequencies, which allows for a more detailed characterization of the relevant flow physics.

Verhulst \& Meneveau used POD analysis of large eddy simulations (LES) to show that large-scale streamwise counter-rotating rolls contribute to the kinetic energy in entrainment of high-velocity wind into the wind turbine array \cite{RN313}. It has also been reported that the energy of POD modes was associated with the characteristic wavelength of the flow between the turbines \cite{RN301, zhang2020characterizing}. Generally, POD modes with high kinetic energy were associated with a large characteristic wavelength in wind farms \cite{RN301}. Zhang \& Stevens \cite{zhang2020characterizing} used POD to study the flow structures within and above large wind farms. The POD modes of different flow patterns have been identified with physically-relevant frequencies. Hamilton {\it et al.}\ \cite{hamilton2018generalized} constructed a reduced-order model for wind turbine wake based on POD modes and showed that their model can accurately predict the turbine wake dynamics in terms of large-scale structures.

Due to the enormous amount of data required for a detailed DMD analysis, previous DMD studies were limited to analyzing the flow around a single or limited number of turbines. Premaratne \& Hu \cite{RN420} used DMD to study the data from experimental particle image velocimetry measurements of the flow behind a wind turbine. Their analysis reveals that vortical structures evolve from axisymmetric to asymmetric shape before the vortices break up. A data-driven approach built on Hankel-based DMD was implemented by Ali \& Cal \cite{RN421} to study the wake characteristics behind a wind turbine array, and then a predictive model was built to forecast the flow evolution on small time scales. The model also captures the chaotic transition between the center of the rotor and the atmospheric streamflow caused by wake advection and vortex shedding. Iungo {\it et al.}\ \cite{RN442} proposed a data-driven reduced-order model to study the wake behind a wind turbine using DMD. They showed that important DMD modes can be connected to, for example, the blade rotation frequency. Annoni {\it et al.}\ \cite{RN425} and Cassamo $\&$ van Wingerden \cite{cassamo2020potential} used a so-called input-output Dynamic mode Decomposition (IO-DMD) method in which the DMD system is extended by considering input and output observables. These studies showed that it is possible to construct a model capable of reconstructing wind turbine wakes quite accurately using a limited number of modes, demonstrating the potential DMD approach for the design of wind farm controllers. In the present study, we will analyze the flow of high-resolution wind farm simulations, considering DMD analysis on a much larger scale than previously considered. The analysis will assess the potential of different DMD modes to reconstruct the flow fields in wind farms using a limited number of modes.

The wind farm layout has a strong influence on the wind farm performance and large-scale flow structures in the wind farms. The work by Lissaman \cite{lissaman1979energy} in 1979 symbolizes the start of wind farm layout optimization, and Mosetti {\it et al.}\ \cite{RN356} pioneered the use of wind farm layout optimization using genetic algorithms. Also, various LES studies (see, e.g.\,  Archer {\it et al.}\ \cite{RN293}, Wu \& Port\'e-Agel \cite{RN360}, and Stevens {\it et al.}\ \cite{RN299}, and many others) have shown that horizontal staggering improves wind farm efficiency while significantly affecting the dominant flow structures. The use of vertical staggering to improve wind farm performance is less well explored. However, it is known from experiments (see e.g.\ Vested {\it et al.}\ \cite{RN283}, Chamorro {\it et al.}\ \cite{chamorro2011turbulent}), LES (see e.g.\ Vasel-Be-Hagh $\&$ Archer \cite{RN287}, Zhang {\it et al.}\ \cite{RN275,RN305}, Wu {\it et al.}\ \cite{wu2019power}), and model calculations (see e.g.\ Wang {\it et al.}\ \cite{RN286}, Chamorro {\it et al.}\ \cite{RN282}, {Stanley {\it et al.}\ \cite{Stanley2019,stanley2019optimization}}) that vertical staggering can improve wind farm performance. {\color{black}Besides, Abkar and Port\'e-Agel \cite{Abkar2015} and Xie and Archer \cite{Xie2015} have shown that the wake expansion in the horizontal direction is faster than in the vertical direction. As the wake recovery is slower in the vertical direction, it is more effective to displace wind turbines in the vertical direction such that they are outside the wake of preceding turbines, although the vertical displacement of turbines is limited by the turbine height that can be achieved.} Therefore, we will perform LES of different wind farm layouts to assess the ability of DMD to capture the flow structures in various cases. We consider four different wind farm layout, exploring the effect of horizontal and vertical staggering, which also reveals that even for large wind farms combining horizontal and vertical staggering is beneficial.

To summarize, previous DMD studies have focused on the flow structure around a single or a limited number of turbines. In contrast, in the present study, we focus on the ability of DMD to capture the essential dynamical modes in extended wind farms, thus providing an assessment of the potential of DMD to develop reduced-order models that are suitable to study wind farm dynamics and wind farm control strategies.\ In Sec.\ \ref{sec2} we introduce the simulation method and DMD analysis techniques. The DMD analysis is provided for four distinctly different wind farm layouts, and in Sec.\ \ref{sec3} we reveal that a combination of horizontal and vertical staggering is beneficial for wind farm power production. In Sec.\ \ref{ssec3_3} the DMD analysis is presented, focusing on the ability of amplitude mode selection (AP method) and sparsity-promoting (SP method) to capture the relevant flow modes and the ability to reconstruct the flow with a limited number of modes meaningfully. The main conclusions are presented in Sec.\ \ref{sec4}.


\section{modeling methodology and analysis methods}\label{sec2}
Essential aspects of LES of wind farms are introduced in subsec.\ \ref{ssec2_1}. The DMD application to characterize wind farm flows is discussed in subsec.\ \ref{ssec2_2}. 

\subsection{LES for wind farm flows}\label{ssec2_1}
We use LES to simulate the flow in large wind farms in a neutral atmospheric boundary layer (ABL). Turbines are modeled using the actuator disk model \cite{RN404} with the assumption that they operate in regime II in which the turbine thrust coefficient $C_T$ is constant \cite{johnson2004methods}. In this work, we assume that all turbines operate at $C_T=0.75$ \cite{RN406}. The governing equations that are solved in the LES are:
\begin{equation}
\begin{aligned}
	\frac{\partial{\tilde{u}_i}}{\partial{t}}+\frac{\partial(\tilde{u}_i\tilde{u}_j)}{\partial{x_j}}&=-\frac{\partial{\tilde{p}^*}}{\partial{x_i}}-\frac{\partial{{\tau}_{ij}}}{\partial{x_j}}-\delta_{i1}\frac{1}{\rho}\frac{\partial p_{\infty}}{\partial x}+f_i\\
	\frac{\partial{\tilde{u}_i}}{\partial{x_i}}&=0
\end{aligned}
\label{eq1}
\end{equation}
where $\tilde{u}_i$ represents the resolved velocities, $f_i$ denotes the forces from the turbines modeled by the actuator disk model, $\tilde{p}^*$ represents the filtered modified pressure written as $\tilde{p}^*=\tilde{p}/\rho+\frac{1}{3}{\tau}_{kk}-p_{\infty}/\rho$. ${\tau}_{ij}$ is the subgrid-scale stress term, whose deviatoric part $({\tau}_{ij}-\frac{1}{3}\tau_{kk}\delta_{ij})$ is modeled by the scale-dependent Lagrangian dynamic model \cite{RN398,RN399}. The molecular viscosity term is neglected as we consider very high Reynolds number flows \cite{RN398}. A pseudospectral discretization is utilized in the horizontal directions combined with a staggered second-order centered finite difference scheme in the vertical direction. On the top of the physical domain, a no stress and no flow-through condition is assumed. At the ground, the Monin-Obukhov similarity theory is employed to calculate the shear stress on the surface \cite{RN397}. Time marching is performed using a second-order accurate Adams-Bashforth scheme. Thermal effects and Coriolis effects are not considered. The concurrent precursor method is used to generate realistic atmospheric inflow conditions \cite{RN400,RN300}. The code \cite{Albertson1999} has been validated against the EPFL wind tunnel experiments \cite{RN384,RN401}, the Horns Rev wind farm measurements \cite{RN300}, and high-resolution wake measurements performed at Delft \cite{RN402}.

\begin{figure}[!t]
	\centering
	\includegraphics[width=0.95\textwidth]{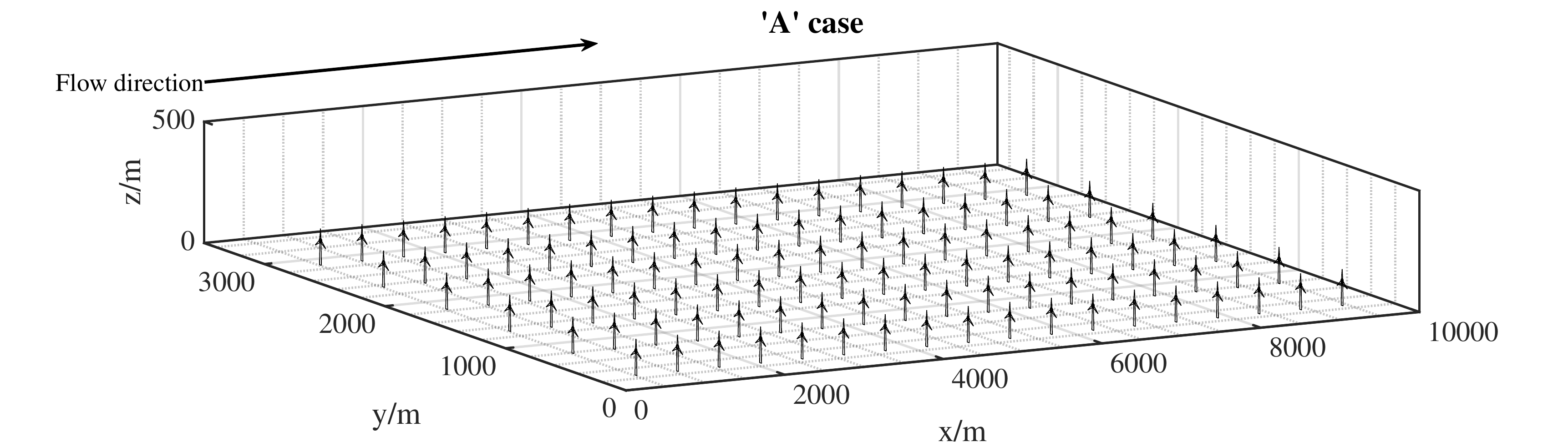}
	\includegraphics[width=0.95\textwidth]{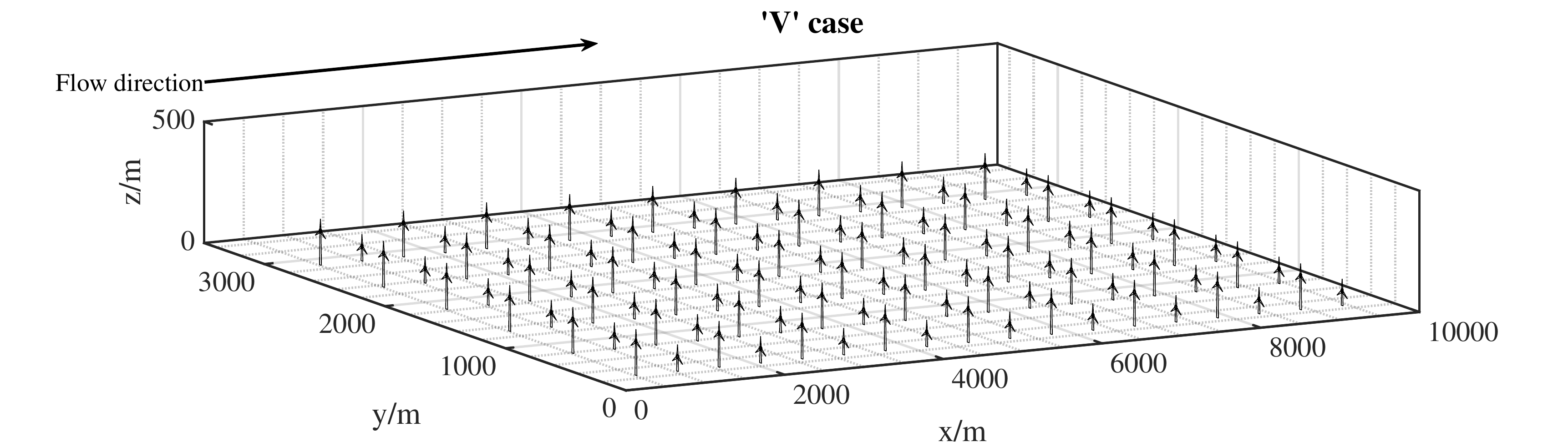}
	\includegraphics[width=0.95\textwidth]{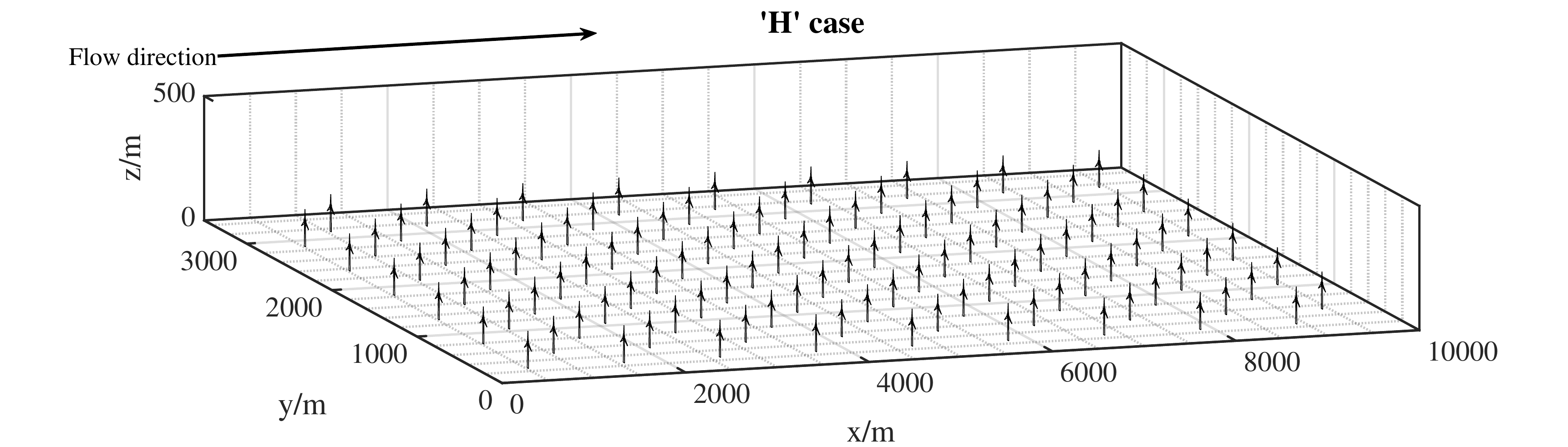}
	\includegraphics[width=0.95\textwidth]{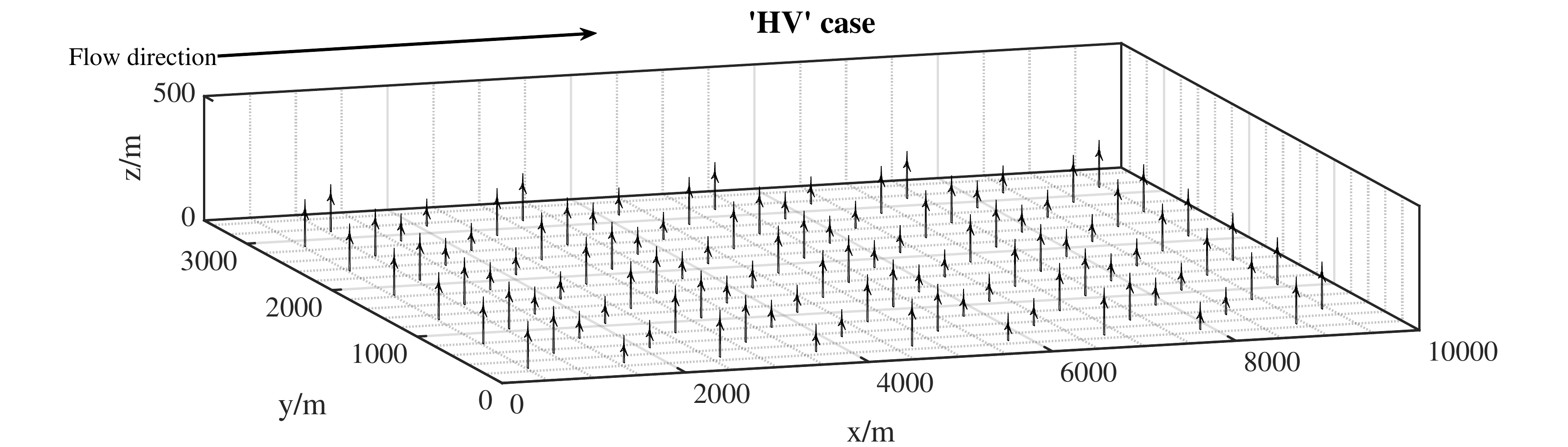}
	\caption{Sketch of the turbine layout for considered cases. }
	\label{fg1}
\end{figure}

To assess the impact of the wind farm layout on the flow structures in the wind farm using DMD we consider four different layouts, namely: (1) Alignment in the horizontal and vertical direction; (2) Alignment in the horizontal direction while staggering in the vertical direction; (3) Alignment in the vertical direction while staggering in horizontal direction; (4) A combination of vertical and horizontal staggering. The four cases are abbreviated as `A', `V', `H', and `HV', see Fig.\ \ref{fg1} for a sketch. The vertical staggering is achieved by elevating the height of turbines in odd rows and lowering the even rows. The horizontal staggering is obtained by shifting the position of turbines in even rows by half of the spanwise turbine spacing in the lateral direction.

\begin{table}[h]	
	\centering
	\caption{Wind farm and turbine parameters}
	\begin{tabular}{p{10cm}l}
		\toprule
		\textbf{Parameters} & \textbf{Values}\\
		\midrule
		Height of domain $H (m)$ & $1000$\\
		Middle hub-height $H_t(m)$ & $100$\\
		Turbine diameters $D_t(m)$ & $100$\\
		hub-height differences $H_d(m)$ & $40$\\
		Length of domain in streamwise direction $L_x$ & $4\pi H$\\
		Length of domain in spanwise direction $L_y$ & $\pi H$\\
		Spanwise turbine spacing $S_y$ & $5.24D_t$\\
		Streamwise turbine spacing $S_x$ & $5.24D_t$\\
		\bottomrule
	\end{tabular}
	\label{tb1}
\end{table}

All simulations are performed. on a $512\times128\times192$ grid for a three-dimensional domain of $4\pi H\times\pi H\times H$ in the streamwise, spanwise, and vertical direction, respectively. The domain height $H=1000\ m$, the turbine diameter is $D_t=100\ m$, and the default turbine height is $H_t=100\ m$. For consistency with our previous works \cite{zhang2020characterizing,RN382} each wind farm has $108$ wind turbines ($N_x=18$ rows in streamwise and $N_y=6$ columns in spanwise direction). Both the spanwise turbine spacing $S_y$ and the streamwise turbine spacing $S_x$ are constant. It is worthy of mention that for the horizontally staggered cases 'H' and 'HV', the effective distance in which the downstream turbines are mainly affected by the upstream wakes is two times the streamwise turbine spacing, e.g.\ the row 3 is now affected by the wakes of row 1 instead of row 2. $H_d$ refers to the difference in hub-height compared to the normal turbine height $H_t=100\ m$. $H_d$ is $40\ m$, which means that the shorter turbines are $60\ m$ and the taller turbine $140\ m$. {\color{black} Based on the data from Ref.\ \cite{RN275} we find that that for the boundary layer under consideration the power output of the tall $140\ m$ and the short turbine $60\ m$ is about 10\% higher than that for two turbines of average height $100\ m$}. Table \ref{tb1} summarizes the simulation parameters. The horizontal computational domain lengths $L_x$ and $L_y$ are normalized by $H$, and streamwise and spanwise turbine spacing, i.e. $S_x$, and $S_y$, are normalized by $D_t$.

\subsection{Sparsity-promoting dynamic mode decomposition}\label{ssec2_2}
DMD is a powerful tool for flow analysis (see Ref. \cite{RN407} among others). Like POD, a DMD analysis results in modes that can be used to describe the flow physics. Each mode evolves in time and provides compact dynamic information underlying the original physical system. These modes can be regarded as a reduced-order representation of the extracted structures from a physical domain sampled in a time sequence that contains the largest contribution to the evolution of the system. DMD enables one to cast complex flows onto subspaces that constitute a dynamic system whose degrees of freedom are reduced. Sparsity-promoting dynamic mode decomposition (SP-DMD) was proposed by Jovanovic {\it et al.}\ \cite{RN382} to achieve a trade-off between the quality of the approximation and the number of required modes. In other words, it provides an objective method for one to determine quantitively at what order the decomposition can be truncated to achieve a given accuracy. We perform three-dimensional SP-DMD on the flow field created in different wind farm layouts in this work. The algorithm is briefly introduced below the following Ref.\ \cite{RN382}.

Firstly, two data matrices originating from the velocity field measured in temporal sequence are formulated $X\in\mathbb{R}^{3N_{xyz}\times N_t}, X'\in\mathbb{R}^{3N_{xyz}\times N_t}$:
\begin{equation}
X=
[\begin{array}{ccccc}
x_1 & x_2 & x_3 & ... & x_{N_t}
\end{array}]_{3N_{xyz}\times N_t}
\label{eq3}
\end{equation}
\begin{equation}
X'=
[\begin{array}{ccccc}
x_2 & x_3 & x_4 & ... & x_{N_t+1}
\end{array}]_{3N_{xyz}\times N_t}
\label{eq4}
\end{equation}
where $N_t$ indicates the number of snapshots; $N_{xyz}$ accounts for the spatial degrees obtained by $N_x\times N_y\times N_z$, which are the number of grid points in the three spatial directions. Therefore, each row vector of $X,X'$ has size $N_t$, and each column of $X,X'$ has a size of $3\times N_{xyz}$, in which the factor 3 indicates that three velocity components are considered. Note that $X'$ is one step forward than $X$. Then, one can find a time-independent matrix operator which is capable of optimally and linearly relating the two matrices in time such that:
\begin{equation}
X'=AX
\label{eq5}
\end{equation}
where the $A$ operator can be obtained by solving the optimization problem:
\begin{equation}
A=\mathop{\arg\min}_{A}=\left\|X'-AX \right\|_F=X'X^{\dagger}
\label{eq6}
\end{equation}
In Eq.\ \ref{eq6}, $\left\| \ \ \right\|_F$ is the Frobenius norm and the $\dagger$ symbol indicates the pseudo-inverse, which is computed as 
\begin{equation}
X^{\dagger}=V\Sigma^{-1} U^*
\label{eq7}
\end{equation}
where $*$ denotes the complex conjugate transpose, and $V$, $\Sigma$, $U$ can be obtained from singular value decomposition:
\begin{equation}
X=U\Sigma V^*
\label{eq8}
\end{equation}
It is necessary to clarify that directly solving Eq.\ \ref{eq7} is computationally intractable since the spatial orders are large. Hence, one typically uses $ X'X=V\Sigma^2V^*$ and conducts the eigenvalue decomposition to obtain $V$ and $\Sigma$ before obtaining $U$ using $U=XV\Sigma^{-1}$. 

Observing Eq.\ \ref{eq5}, reveals that $A$ tends to have a large dimension. To make the computation more tractable, we consider only the first $r$ leading orders by projecting the $A$ matrix onto $U$. We obtain:
\begin{equation}
F_{DMD}=U^* AU=U^*X'V\Sigma^{-1}
\label{eq9}
\end{equation}
where the $F_{DMD}$ matrix is the optimal reduced representation of the $A$ matrix in the $U$ space.

In the r-dimensional subspace, we should formally have $\bar{x}_t=F_{DMD}\bar{x}_{t-1}$, where the reduced-order $\bar{x}_t$ is related to the original state vector by $x_t=U\bar{x}_t=UF_{DMD}\bar{x}_{t-1}$. $F_{DMD}$ can be expressed in a diagonal coordinate with respect to its eigenvectors $\{\begin{array}{ccc}y_1 & ... & y_r\end{array}\}$ and the corresponding eigenvalues$\{\begin{array}{ccc}\mu_1 & ... & \mu_r\end{array}\}$:
\begin{small}
	\begin{equation}
	F_{DMD}=
	\left[ 
	\begin{array}{ccc}
	y_1 & ... & y_r
	\end{array}
	\right]
	\left[
	\begin{array}{ccc}
	\mu_1 & \cdots & 0\\
	\vdots & \ddots & \vdots\\
	0 & \cdots & \mu_r\\
	\end{array}
	\right]
	\left[
	\begin{array}{c}
	z^*_1\\
	\vdots\\
	z^*_r
	\end{array}
	\right]
	=YD_\mu Z^*
	\label{eq10}
	\end{equation}
\end{small}
where $Y$ and $Z$ are unitary matrices with orthonormal columns, i.e. $z_j^*y_i=\delta_{ji}$. $\phi_i=Uy_i$ indicates the DMD mode, and we can reconstruct the original state as:
\begin{equation}
x_t=UYD_\mu^tZ^*\bar{x}_0=\sum_{i=1}^{r}\phi_i\mu_i^tz_i^*\bar{x}_0=\sum_{i=1}^{r}\phi_i\mu_i^t\alpha_i
\label{eq11}
\end{equation}
where $\alpha_i=z_i^*\bar{x}_0$ can be considered as the amplitude of each DMD mode. Rewriting the above equation in a matrix form gives: 
\begin{equation}
\begin{aligned}
\left[\begin{matrix}
x_1 & x_2 & \cdots & x_{N_t}
\end{matrix}\right]&\approx
\left[\begin{matrix}
\phi_1 & \phi_2 \cdots & \phi_r
\end{matrix}\right]
\left[\begin{matrix}
\alpha_1\\
&\alpha_2\\
&&\ddots\\
&&&\alpha_r
\end{matrix}\right]
\left[\begin{matrix}
1 & \mu_1 & \cdots & \mu_1^{N_t-1}\\
1 & \mu_2 & \cdots & \mu_2^{N_t-1}\\
\vdots & \vdots & \ddots & \vdots\\
1 & \mu_r & \cdots & \mu_r^{N_t-1}\\
\end{matrix}\right]
=\Phi D_\alpha V_{and}
\label{eq12}
\end{aligned}
\end{equation}
where $V_{and}$ is the Vandermonde matrix of the eigenvalues of matrix $F_{DMD}$ . The determination of amplitude vector $\alpha$ is equivalent to solving the following optimization problem:
\begin{equation}
\min_{\alpha} \left\|X-\Phi D_{\mu}V_{and} \right\|_F
\label{eq13}
\end{equation}

In order to have a controllable trade-off between the number of modes and the quality of the approximation, the sparsity structure of amplitude vector is modified accordingly. The optimization problem can be rearranged as:
\begin{equation}
\min_{\alpha}\ \ \ \ \left\|\Sigma V^*-YD_\alpha V_{and}\right\|^2_F+\gamma \sum_{i=1}^r |\alpha_i |
\label{eq14}
\end{equation}
where $J(\alpha)$ is obtained by plugging Eq.\ \ref{eq8} and $\Phi=UY$ into Eq.\ \ref{eq13}, the second term calculates the $L_1$ norm of $\alpha_i$ facilitating the convex optimization tools and $\gamma$ is a user-defined positive regularization parameter.

The above introduces the specific procedures of SP-DMD. We will consider two different mode selection criteria. This first is known as the AP method, which selects DMD modes based on the amplitude of $\alpha_i$ in Eq.\ \ref{eq11}. Specifically, the DMD modes are rearranged based on their amplitude from the highest to the lowest amplitude. Order reduction is achieved by excluding the lower amplitude modes, while the modes with larger amplitudes are retained. The second mode selection criterion we use is the SP method. As introduced above, one can firstly raise the user-defined $\gamma$ to penalize the non-zero elements in the amplitude vector by solving a convex optimization problem. In other words, larger $\gamma$ results in more zero elements in the amplitude vector, which increases its sparsity. The global solution of the convex optimization problem can be solved by the alternating direction method of the multipliers algorithm. The remainder of the non-zero elements is adjusted by optimization to approximate the original space as closely as possible. As such, one can iterate the above process by inputting a set of $\gamma$ ranging from low to high values and obtain a set of amplitude vectors ranging {from a small number of zero elements to a large number of zero elements}. However, it should be noted that the non-zero elements in different amplitude vectors might be different because after the sparsity is increased, the values of the remaining non-zero elements are optimized. For more details, see Ref. \cite{RN382}.
 
\begin{figure}[!t]
	\centering
	\includegraphics[width=0.4\columnwidth]{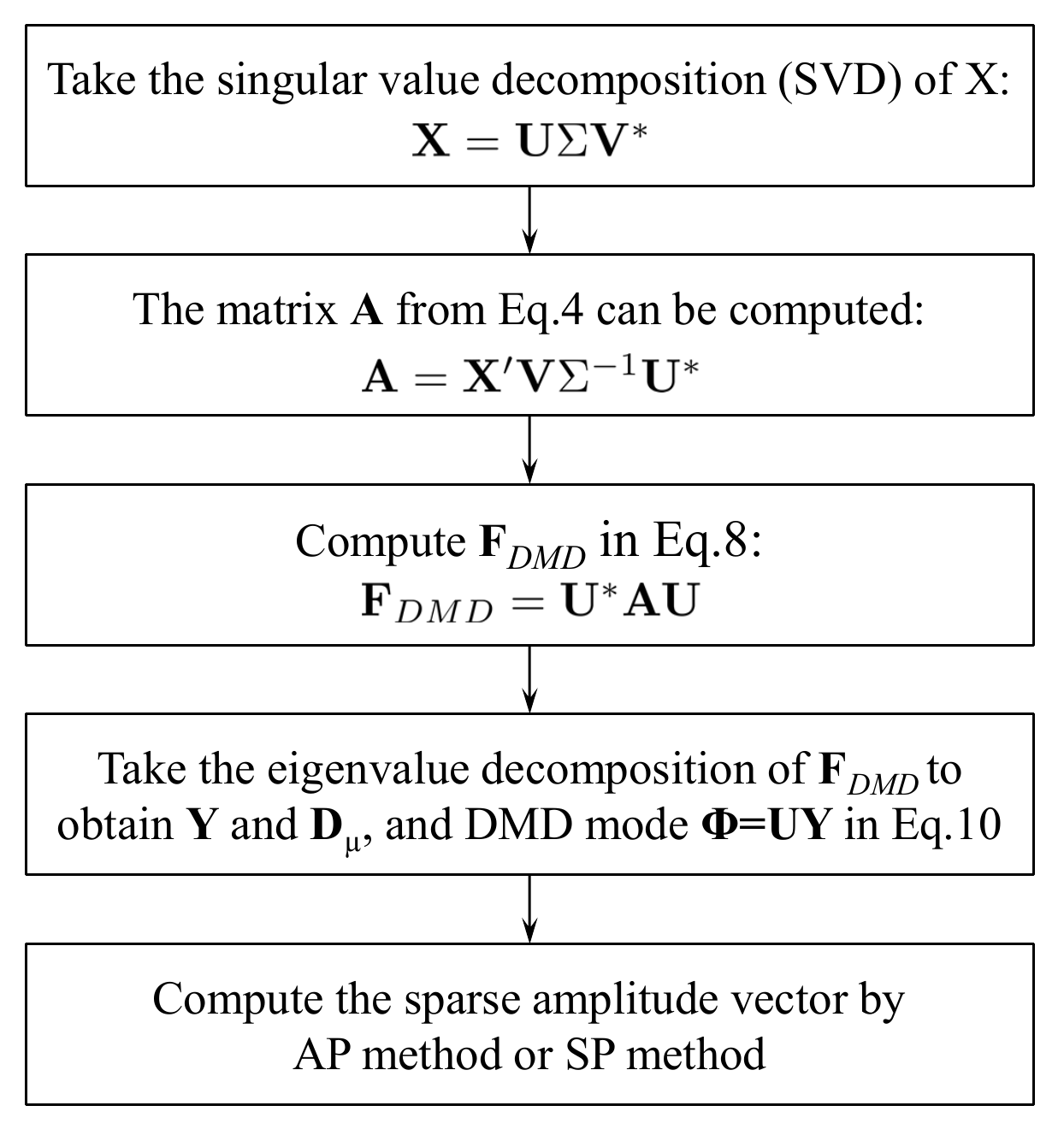}
	\caption{Flow chart of the DMD analysis}
	\label{FlowChart}
\end{figure}

{\color{black} Fig.\ref{FlowChart} briefly summarizes the flow chart of the DMD analysis. Firstly, singular value decomposition (SVD) is implemented on raw data ${X}$, where ${U}$ is the POD mode and $\Sigma$ is the eigenvalue of POD mode, see Eq.\ref{eq8}. Secondly, the mapping operator ${A}$ is computed as shown in Eq.\ref{eq6} and Eq.\ref{eq7}. It is further projected on POD modes to obtain a lower-order representation ${F}_{DMD}$ indicated in Eq.\ref{eq9}. Next, eigenvalue decomposition is implemented on ${F}_{DMD}$ to obtain eigenvalue and corresponding DMD mode denoted as ${D}_{\mu}$ and $\Phi$ respectively, see Eq.\ref{eq10} to Eq.\ref{eq12}. Finally, sparse amplitude vector is calculated based on AP method or SP method.}

\section{Flow and performance analysis for different wind farm layouts}\label{sec3}
In subsec.\ \ref{ssec3_1} we will first discuss the average flow structures for the various wind farm layouts before discussing the wind farm power production of the various layouts in subsec.\ \ref{ssec3_2}.

\subsection{Flow structures in different wind farm layouts}\label{ssec3_1}
\begin{figure}[!t]
	\centering
	\includegraphics[width=0.6\columnwidth]{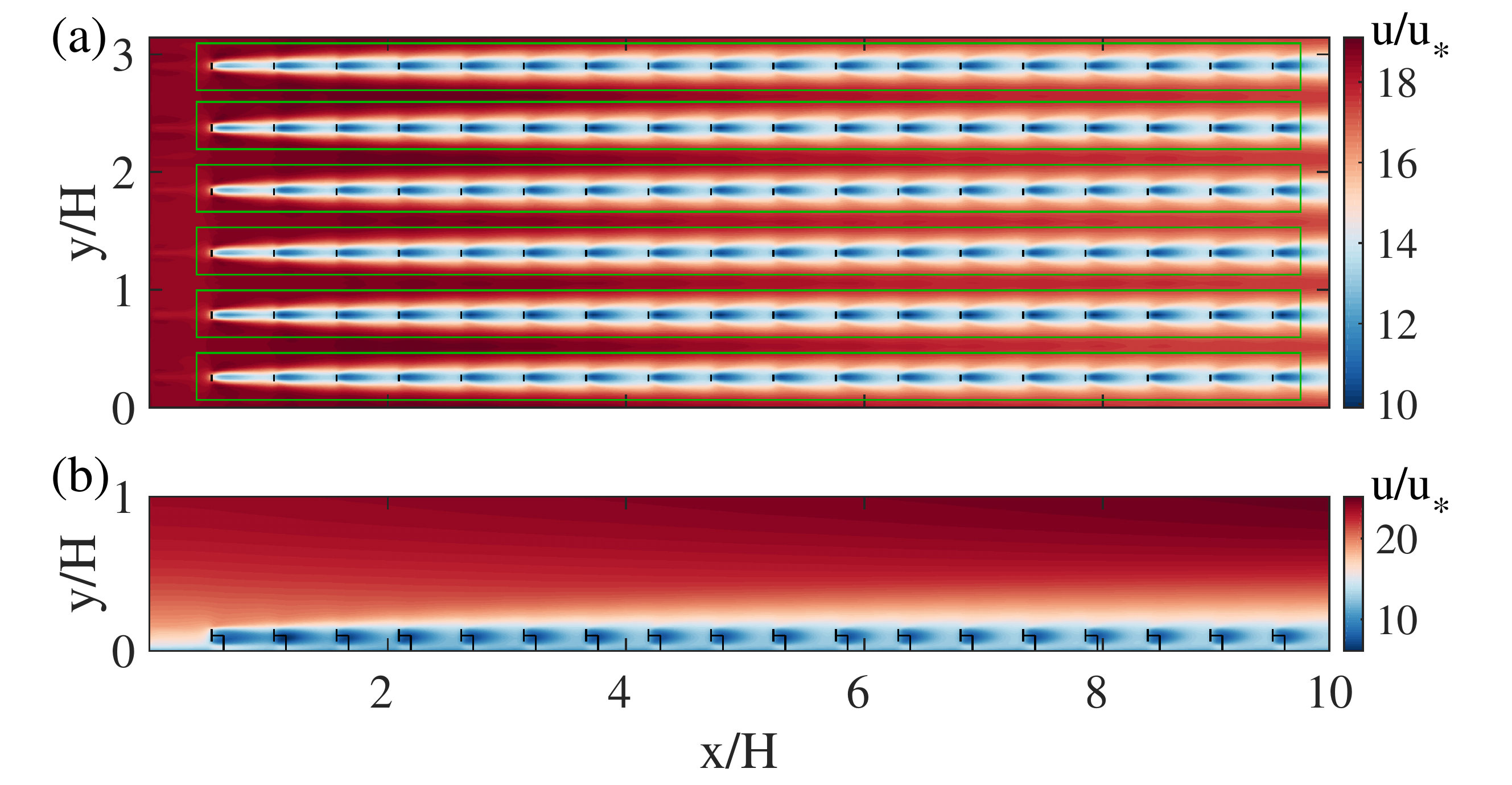}
	\caption{`A' case: (a) streamwise velocity in XY-plane at the turbine hub-height $z/H = 0.1$; (b) streamwise velocity profile in XZ-plane at $y/D_t = 1.31$. Green boxes indicate the regions to compute the averaged velocity of a fluid parcel traveling through turbines.}
	\label{fg2}
\end{figure}

\begin{figure}[!t]
	\centering
	\includegraphics[width=0.6\columnwidth]{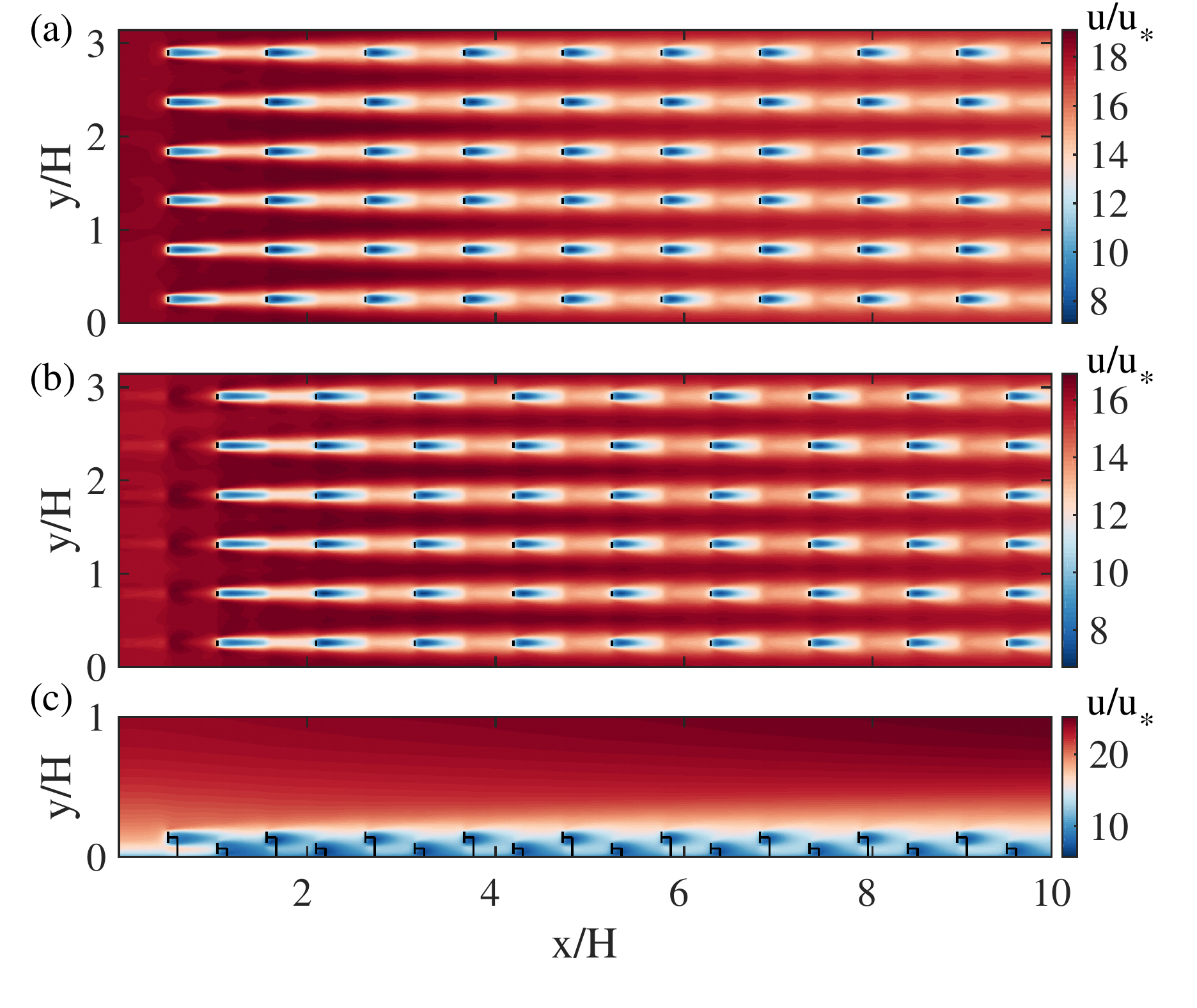}
	\caption{`V' case: streamwise velocity in XY-plane at (a) $z/H = 0.14$, hub-height for tall turbines, and (b) $z/H = 0.06$, hub-height for short turbines; streamwise velocity profile in XZ-plane at $y/D_t = 1.31$.}
	\label{fg4}
\end{figure}

In the `A' case shown in {Fig.\ \ref{fg2}} turbines are aligned in the horizontal and vertical direction. The figure shows the dominant streamwise velocity component of the flow, revealing the wake interaction in the streamwise direction. Clearly, the turbines experience strong wake effects in this layout as they are fully submerged in the wake of relatively close upstream turbines. {\color{black}$u_*=0.45\ m/s$ is the friction velocity} and has been used to nondimensionalise the velocity in our simulations to get typical dimensional results that may be observed in a wind farm (in the sections below, dimensional velocities and frequencies will be shown and discussed. They can be nondimensionalised by this velocity scale and the relevant length scale). 
{Fig.\ \ref{fg4}} shows the streamwise velocity profiles for the vertically staggered case 'V'. Note that, in this case, the turbines are horizontally aligned and vertically staggered, so two streamwise velocity profiles are sketched from different-height XY-planes (at $z/H = 0.14$ and $z/H = 0.06$). Turbines in odd rows are elevated to $140\ m$, and turbines in even rows are lowered to $60\ m$. Comparing {Fig.\ \ref{fg2}}(b) with {Fig.\ \ref{fg4}}(c) reveals that the flow in the vertical direction is more complex than for the `H' case as the vertical staggering leads to a zigzag pattern in the average velocity field. Comparison between {Fig.\ \ref{fg2}}(a) and {Fig.\ \ref{fg4}}(a)(b) suggests that the wake effects in the `V' case are smaller than for the `H' case. This is confirmed in Fig.\ \ref{fg4}(c), which shows that the difference in hub-height between the short and tall turbines limits the impact of the wakes on the performance of downstream turbines.

\begin{figure}[!t]
	\centering
	\includegraphics[width=0.64\columnwidth]{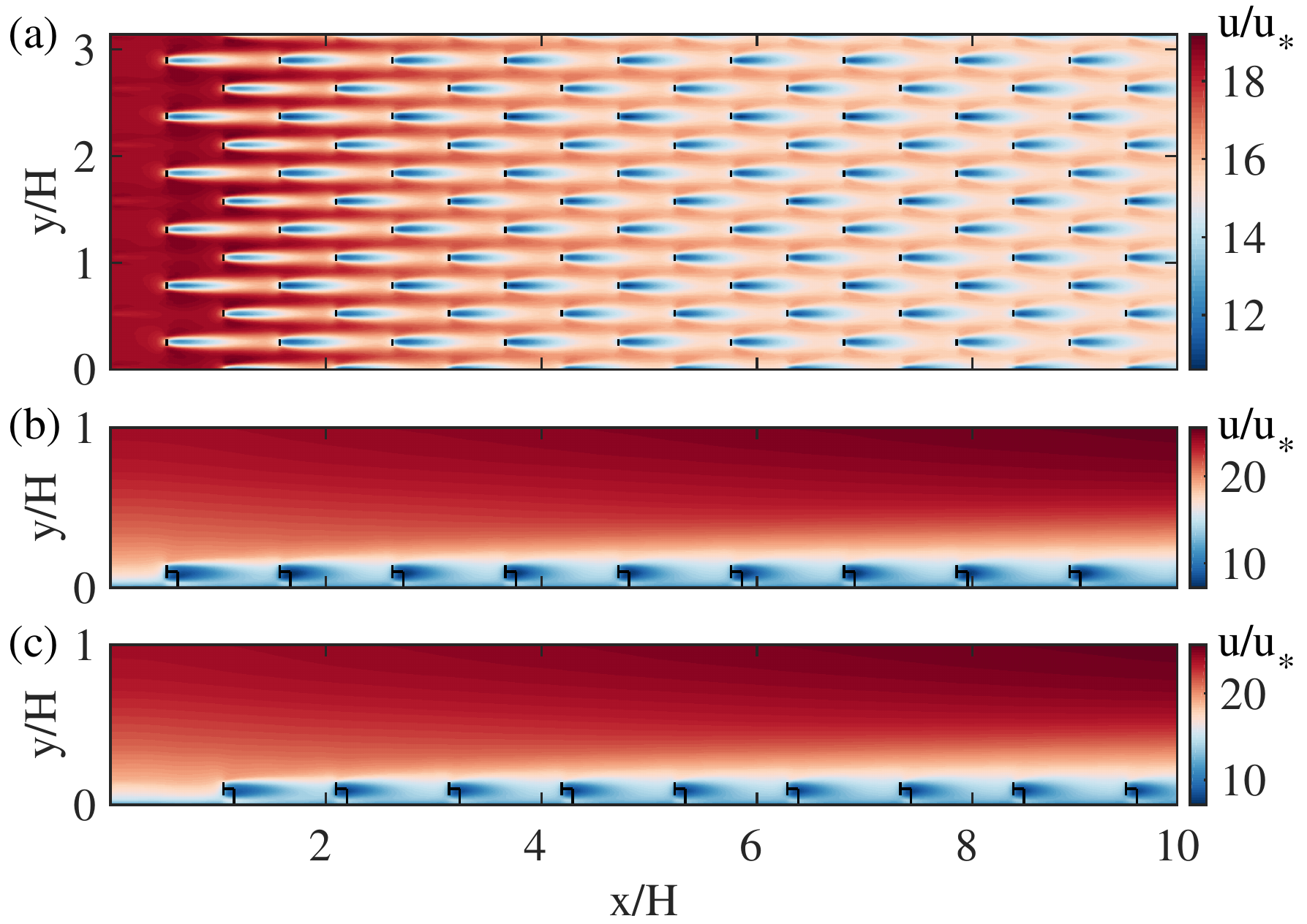}
	\caption{`H' case: (a) streamwise velocity profile in XY-plane at the turbine hub-height $z/H = 0.1$; streamwise velocity profile in XZ-plane at spanwise position of (b) $y/D_t = 1.57$ (c) $y/D_t = 1.31$.}
	\label{fg6}
\end{figure}

\begin{figure}[!t]
	\centering
	\includegraphics[width=0.64\columnwidth]{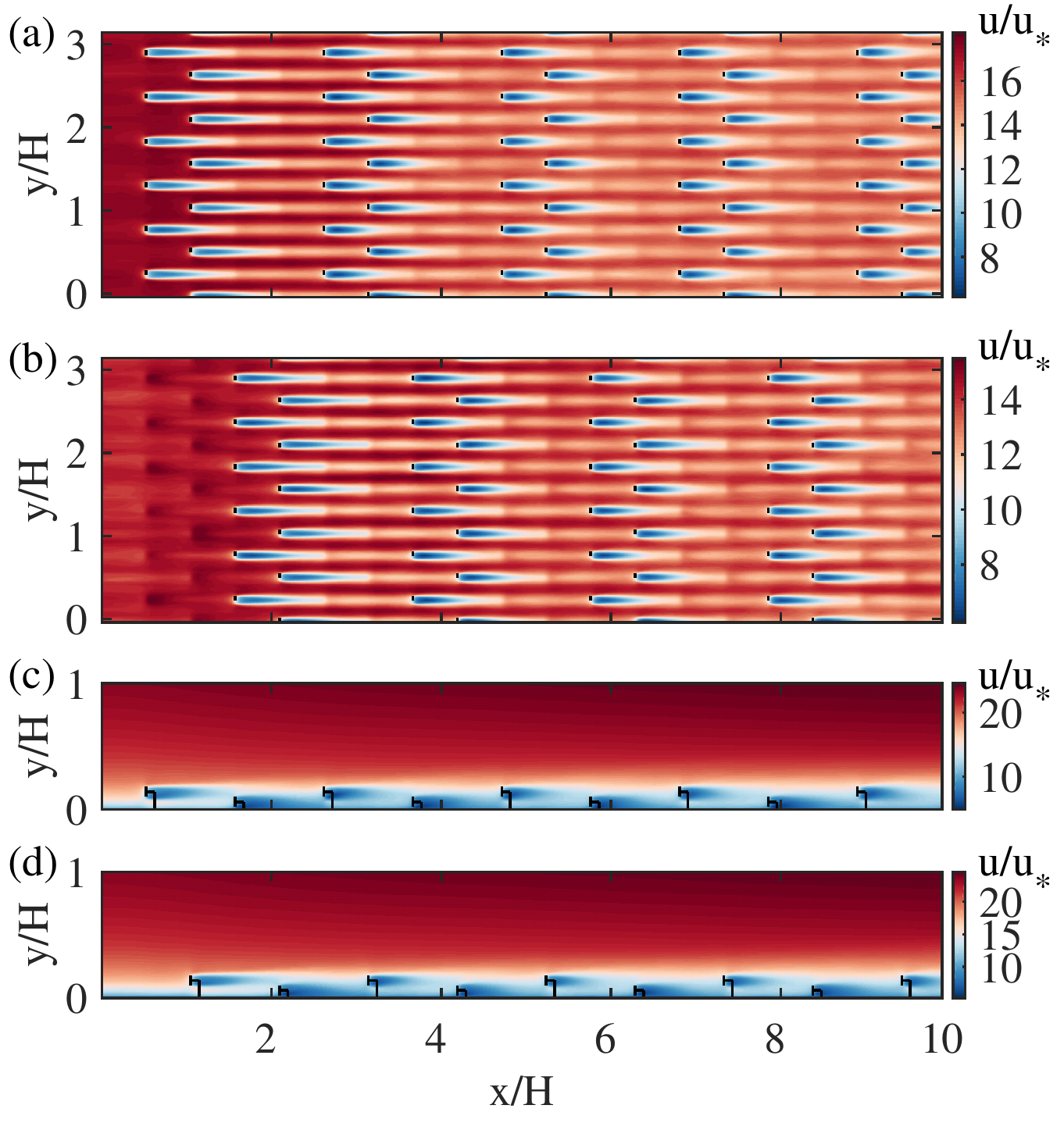}
	\caption{`HV' case: streamwise velocity in XY-plane at (a) $z/H = 0.14$, hub-height for tall turbines and (b) $z/H = 0.06$, hub-height for short turbines; streamwise velocity profile in XZ-plane at spanwise direction of (c) $y/D_t=1.57$ (d) $y/ D_t =1.31$.}
	\label{fg8}
\end{figure}

{Fig.\ \ref{fg6}} shows the streamwise velocity in different planes for the horizontally staggered case `H'. The visualization shows that the wakes from upstream turbines have significantly recovered before reaching the next turbine. However, the velocity in the entire wind farm area is now affected by wakes, whereas the results for the `A' case revealed the presence of high-velocity wind speed channels in between the turbines see also e.g.\ Wu \& Port\'e-Agel \cite{RN360} and Stevens {\it et al.}\ \cite{RN299}. {Fig.\ \ref{fg8}} presents the streamwise velocity profiles for the case `HV'. Since, in this case, the turbines are horizontally and vertically staggered we present the average flow field at two different heights (panels a,b) and two different spanwise locations (panels c,d). The figure suggests that the wake effects are the smallest for this configuration as the velocity just in front of the turbines appears stronger than for the other wind farm layouts. This impression will be confirmed in the next section, in which we analyze the power production of the various cases.

\subsection{Effect of layout on power production}\label{ssec3_2}

\begin{figure}[!t]
	\centering
	\includegraphics[width=0.48\textwidth]{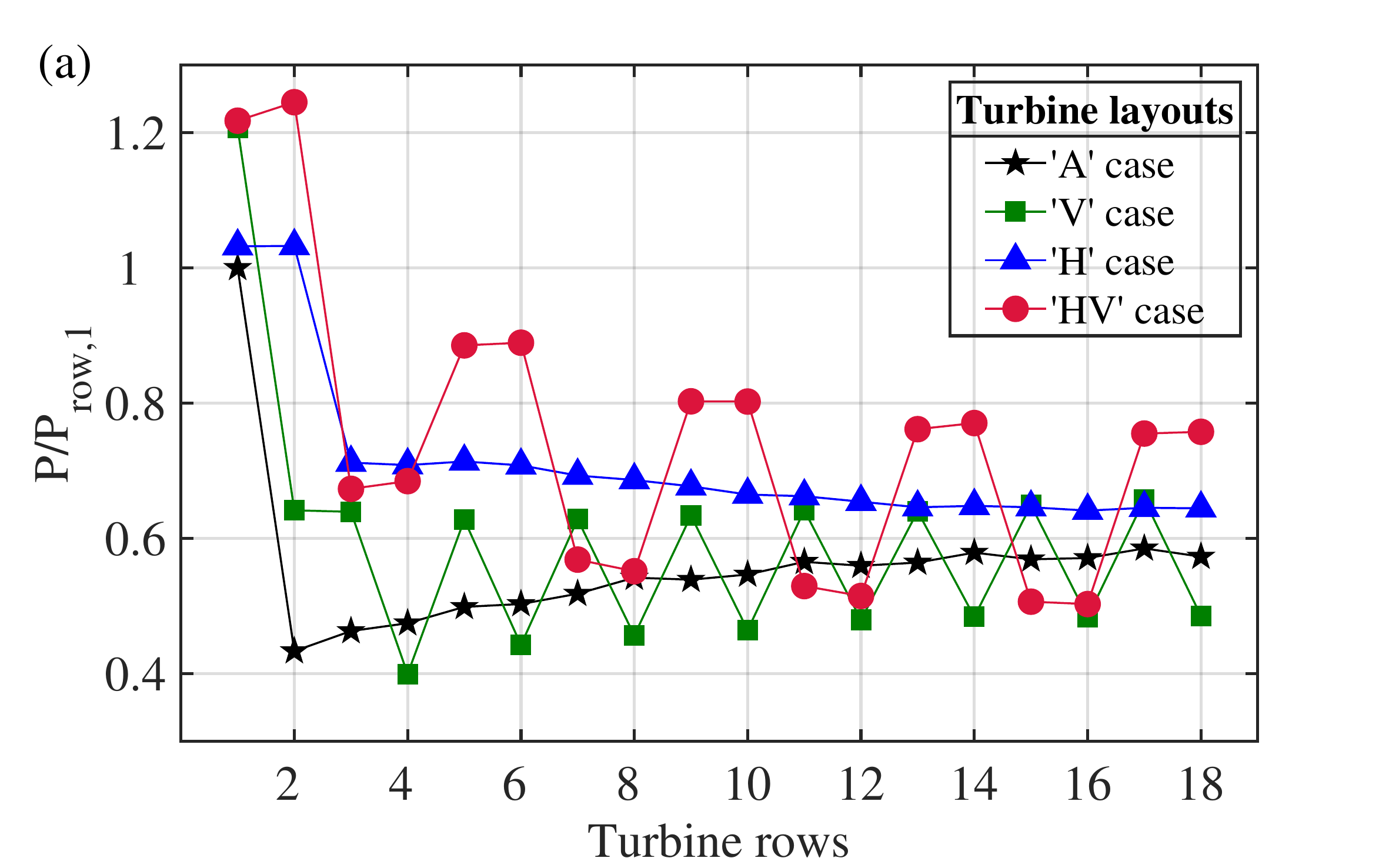}
	\includegraphics[width=0.48\textwidth]{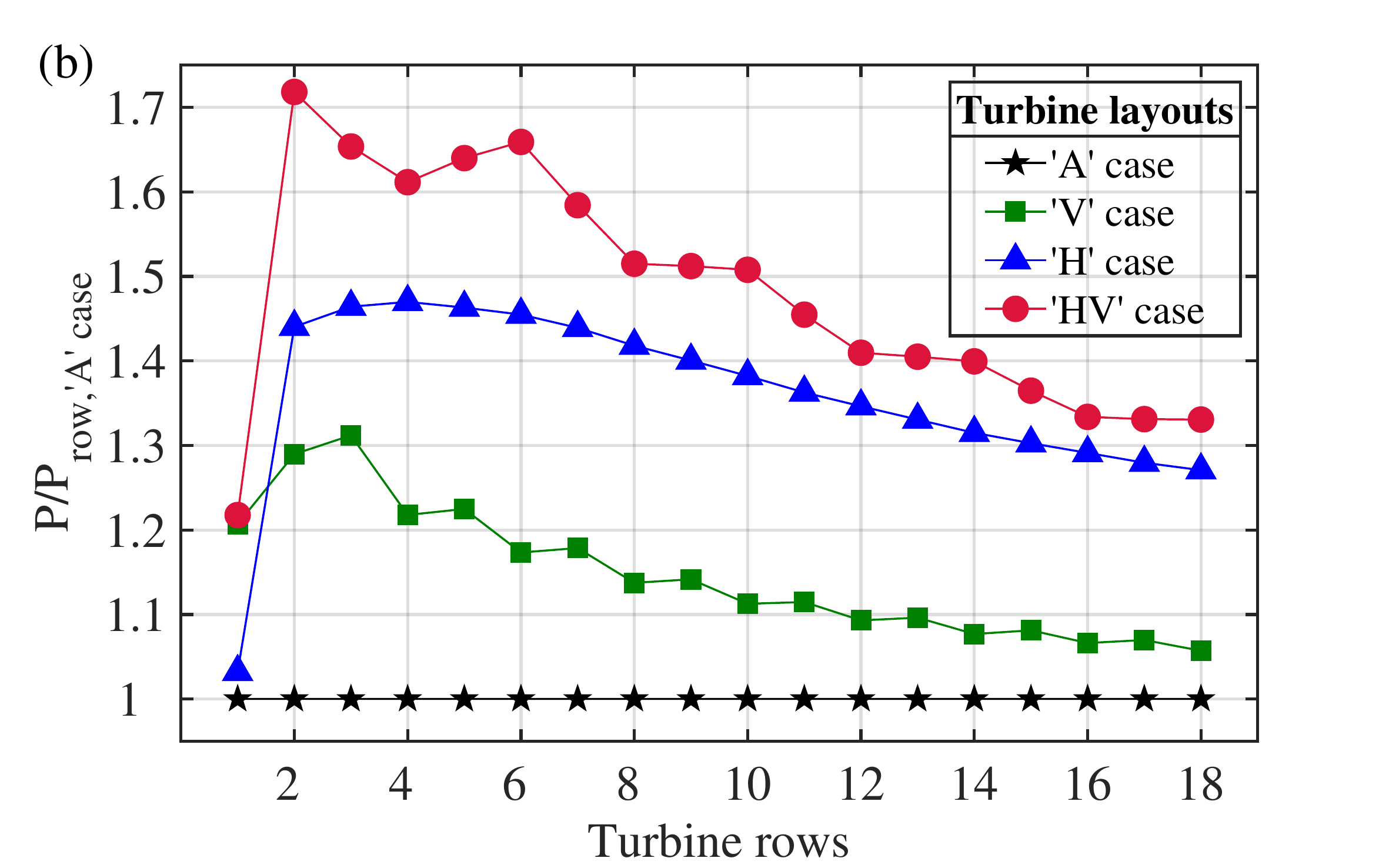}
	\caption{(a) Normalized power production of each turbine row (b) Normalized cumulative power production as function of the downstream direction.}
	\label{fg10}
\end{figure}
{\color{black} To study the effects of the wind farm layout, we calculate the time-average power output as $P=-\left<FU_d\right>$, where $\left<\right>$ denotes time averaging. $F=-\frac{1}{2} C'_T\rho U_d^2A$ accounts for the local force used in turbine model. Here, $U_d$ stands for the disk-average velocity, and $A=\pi D_t^2/4$ is the turbine area. Plugging the expression of $F$ into that of $P$, we can write $P=\frac{1}{2} C'_T\rho \left<U_d^3\right>A$, see Refs. \cite{RN305,RN275,meyers2010large,RN299} for more details.} 
In general, both horizontal and vertical staggering can be used to ensure that wake effects are limited. For the horizontally staggered layout, the downstream turbines are also located outside the wake of turbines in the preceding row.

{Fig.\ \ref{fg10}}(a) presents the power production of each turbine row. Note that the power output is normalized by that of the first row of the corresponding case. We find that vertical staggering (case `V' and case `HV') exhibit a zigzag shape due to logarithmic ABL profile and the vertical staggering, while the curves of the case `A' and case `H' develop smoothly. {Fig.\ \ref{fg10}}(b) presents the cumulative power as a function of downstream turbine row, normalized with the reference aligned case `A'. The figure shows that vertical staggering case `V' performs significantly better than the reference case `A' in the entrance region. However, the relative benefit of vertical staggering decreases further downstream as the `V' case gradually asymptotes to the result of case `A'. The results are in good agreement with that of Zhang {\it et al.}\ \cite{RN305,RN275}. The reason is that in the entrance region, the taller turbines can benefit from the stronger winds at higher elevations. However, in the fully developed region, the vertical kinetic energy transfer is the main energy transfer mechanism, and this mechanism does not differ much in the fully developed region between `A' and `V' cases \cite{RN291,RN299,RN284}.

The figure reveals that the horizontally staggered layout `H' is more effective in increasing the wind farm power production than the `V' case. The reason is that in the horizontal direction there is more space than in the vertical direction to distribute the turbines such that wake effects are limited. Interesting, when horizontal and vertical are combined, i.e. the `HV' case, it leads to a significantly better wind farm performance than using just horizontal or vertical staggering alone. The figure shows that also in the fully developed region the `HV' case is beneficial over the `H' case. Even though this is not our primary focus, it is interesting to point out that the 'HV' case layout outperforms all the other wind farm configurations. This signifies that combining the horizontal and vertical staggering (see Fig.\ \ref{fg1}) can further improve wind farm performance.

\section{Flow characterization by DMD}\label{ssec3_3}
In subsec.\ \ref{ssec3_31} we discuss the convergence criteria and the required sampling frequency for the DMD analysis. In subsec.\ \ref{ssec3_32} we analyze the flow structures obtained using the AP and SP method, and in subsec.\ \ref{ssec3_33} we discuss the accuracy of the flow reconstruction that can be obtained using both methods.

\subsection{Convergence and frequency selection} \label{ssec3_31}
This section presents three-dimensional DMD analyses of the flow structures for the four aforementioned wind farm layouts. {\color{black} It is worthy mentioning that in our DMD analysis (following Ref. \cite{RN407}), the POD modes are first computed. As shown in Fig.\ref{FlowChart}, the first step to implement SVD on raw data gives the POD modes and the corresponding eigenvalues, i.e. ${U}$ and $\Sigma$ respectively (see Eq.\ref{eq8}). In order to obtain converged POD modes, it is necessary to determine the number of required flow snapshots by checking the convergence of eigenvalues corresponding to POD modes.} To verify this we follow the approach by Zhang \& Stevens \cite{zhang2020characterizing} and Newman {\it et al.}\ \cite{RN301}. We calculate the L2 norm of the normalized eigenvalue spectrum as follows.
\begin{equation}
\centering
\eta^{N_t}=\left\|\frac{\lambda_i}{\sum_{i=1}^{N_t}\lambda_i}\right\|^2
\label{eq16}
\end{equation}
where $\lambda_i(1\leq i \leq N_t)$ represents the eigenvalues of corresponding POD modes in ${U}$ {\color{black}(not to be confused with the eigenvalue in Eq. \ref{eq10}}). The relative error is then given by:
\begin{equation}
\centering
E(N_t)=\left|1-\frac{\eta^{N_t}}{\eta^\infty} \right|
\label{eq17}
\end{equation}
where the reference value $\eta^{\infty}$ is determined using $N_t=5000$ snapshots. The number of tested snapshots is 50, 100, 250, 500, 625, 1000, 1250, 2500, and 5000, respectively. The snapshots are selected from the available simulation length, which means that the sampling frequency is higher when more snapshots are used.

\begin{figure}[!t]
	\centering
	\subfigure{
		\includegraphics[width=0.31\textwidth]{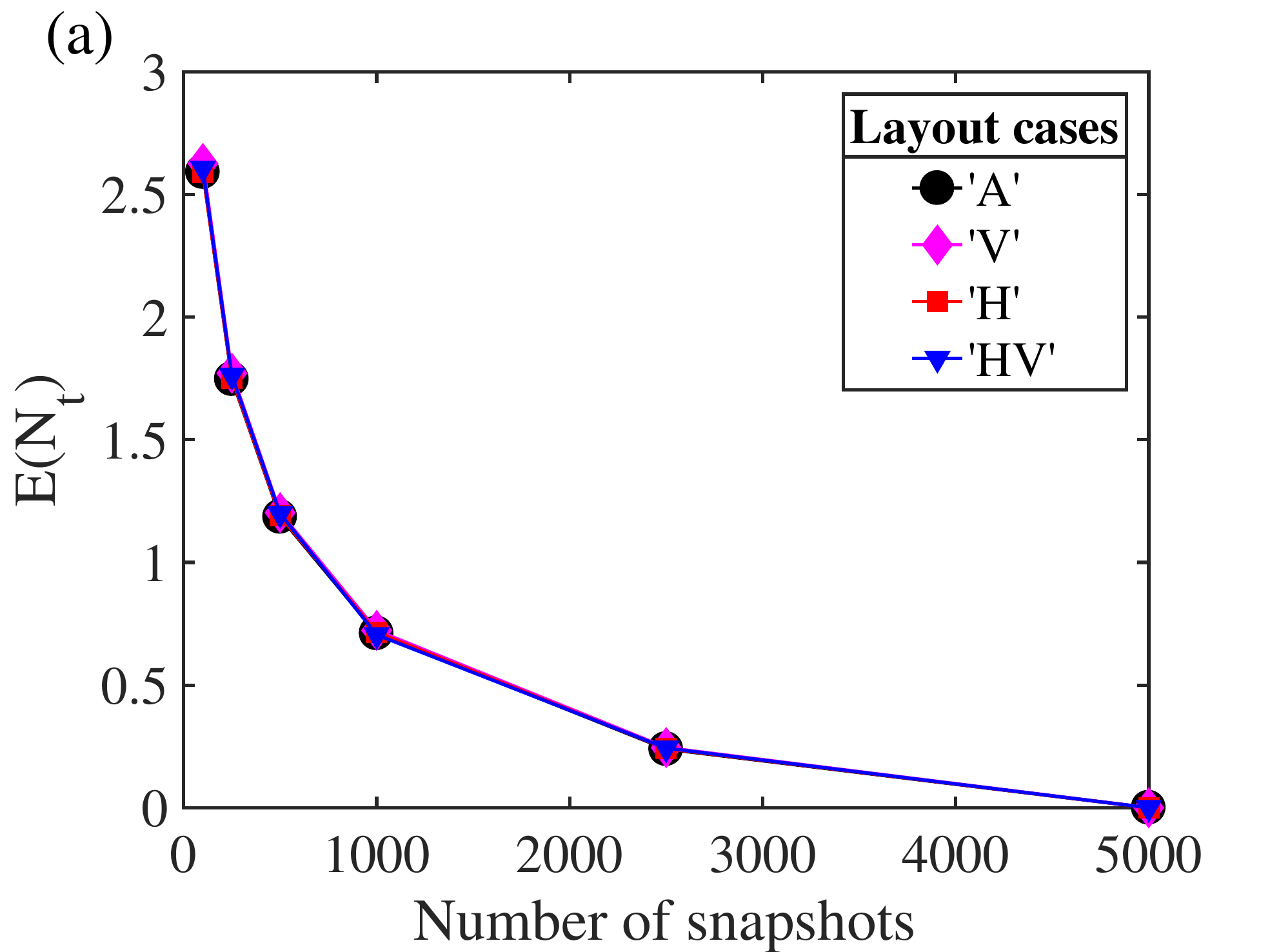}
	}
	\subfigure{
		\includegraphics[width=0.31\textwidth]{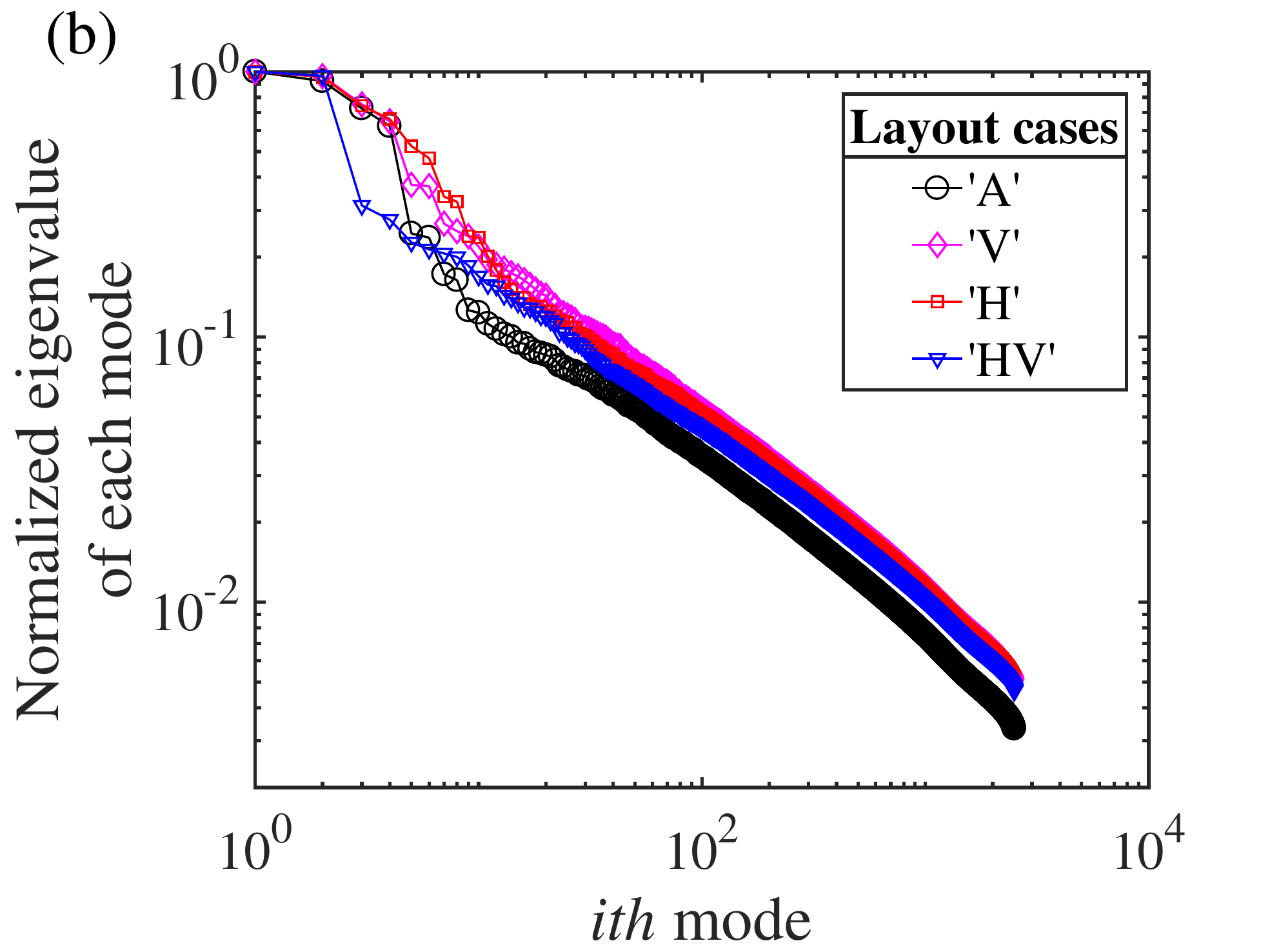}
	}
	\subfigure{
		\includegraphics[width=0.31\textwidth]{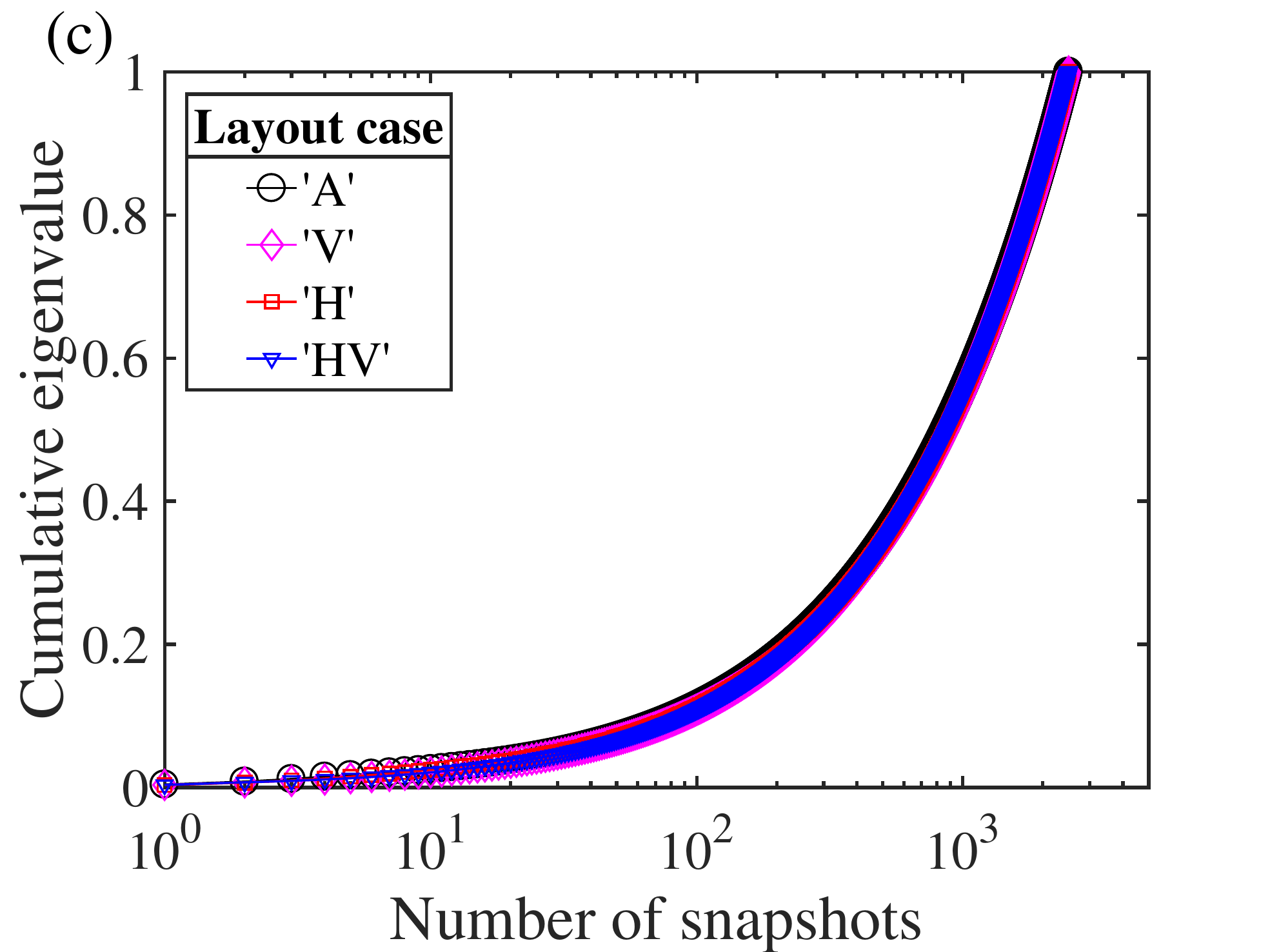}
	}
	\caption{(a) Convergence test of the POD results for different number of snapshots (b) Eigenvalues of each POD mode for 2500 snapshots (c) Cumulative POD eigenvalues for 2500 snapshots}
	\label{fg11}
\end{figure}

The result of the convergence analysis is presented in Fig.\ \ref{fg11}. Panel(a) demonstrates that $E(N_t)$ gradually decreases when the number of snapshots is increased and reaches a relatively small value at 2500 snapshots. The eigenvalues in panel (b) can be interpreted as the kinetic energy in the POD modes \cite{Berkooz1993}. Thus, similarly to our previous works \cite{zhang2020characterizing}, we can see that some POD modes (lower-ranking in $i$) are associated with higher kinetic energy than higher-ranking modes. Panel (c) shows the cumulative eigenvalues, which reveals that about 1000 POD modes are required to capture 60\% of the kinetic energy.
 
{\color{black} In addition to determining the necessary amount of snapshots to obtain converged POD modes, we also need to ensure that we capture the dynamically important flow physics.} Therefore, we compare the sampling frequency ($f_s$) with the frequency corresponding to the average travel time of a fluid parcel between two turbines ($f_t$) in the streamwise direction. Note that both of them have physical unit $Hz$. The following shows how to compute them. 
{To compute $f_t$, we first obtain the mean velocity of a fluid parcel traveling through the turbine spacing at the turbine height by taking the spatial average of the velocity in the regions encompassed by the green boxes as shown in Fig.\ \ref{fg2} (the conclusion below is not sensitive to the actual size of the boxes). The overall mean velocity in these regions is calculated to be around $6.5m/s$. Consequently, we can have $f_t= 6.5\ m/s\div524\ m\approx0.012\ Hz$ ($524 \ m$ is the streamwise turbine distance in the aligned case). This frequency is halved for the cases in which the turbines are horizontally staggered. 
$f_s$ is selected based on the Nyquist criterion that the lower bound sampling frequency by which the process can be identified is at least twice as high as its inherent frequency. In principle, more snapshots in a fixed span of the flow data acquires more dynamic information with a higher frequency. However, this would be computationally expensive. {\color{black}Thus, as a compromise, we decided to use 2500 snapshots separated by 150 time steps for the DMD analysis below, which are sufficient to provide converged POD modes as mentioned above.} This results in a sampling frequency $f_{s}=1\div(0.02222\ s\times150)\approx0.3\ Hz$ where $0.02222s$ is the time step in seconds in our simulations (we use 1000 m as the length scale, 0.45 m/s as the velocity scale and $10^{-5}$ time units as the nondimensional time step). $f_{s}$ is thus about 25 times higher than $f_t$, which is high enough to capture the relevant flow physics in our wind farms. }In the remainder of the manuscript, we will present the DMD results analysis, focusing on the difference between the AP and SP method.

\subsection{Comparison of AP and SP method} \label{ssec3_32}
\begin{figure}[!t]
	\centering
	\includegraphics[width=1\textwidth]{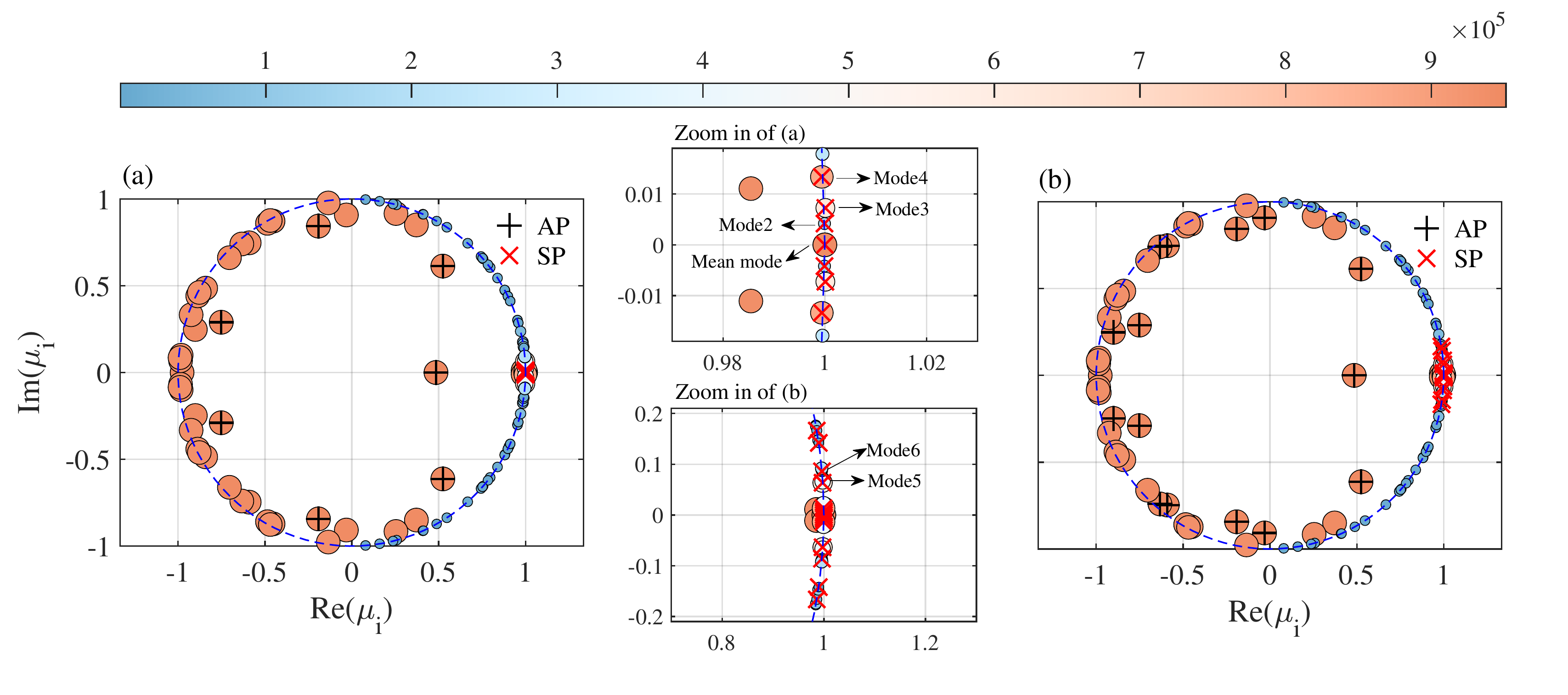}	
	\caption{ {The 100 representative DMD modes expressed by circles with colors. Circles superposed by black plus: modes selected by the AP method; Circles superposed by the red cross: modes selected by the SP method. (a) the first 7 modes selected by the two methods respectively; (b) the first 15 modes selected by the two methods respectively. The color and the size of each mode indicate its amplitude. {\color{black} The {\it mode numbers} indicated in the zoom-in panels correspond to the leading order modes determined using SP method presented in Fig.\ref{fg13}(c).}}}
	\label{fg12}
\end{figure}

Fig.\ \ref{fg12} displays the DMD modes selected by the AP and SP methods. As it is impossible to distinguish all 2500 DMD, we only present the 100 representative modes here. According to the color bar, both the size and the color of the circles indicate the amplitude of corresponding DMD modes; see Eq.\ \ref{eq11}. We have also used black plus and red cross signs to highlight the first DMD modes selected by the AP and SP methods, respectively. Panels (a) and (b) show the first 7 and 15 DMD modes selected by the two methods, respectively. One can immediately notice that the modes selected by the two methods are different. 

{\color{black} Recall in subsec. 2.2 that the AP modes are selected based on the magnitude of the amplitude vector of corresponding element $\alpha_i$ (see Eq.\ref{eq11}) in the amplitude vector. The elements with higher magnitude in the amplitude vector will be chosen in priority.} The AP method reflects a tactic of determining the modes that have strong influences on a system's response resulting from the typical initial condition in the snapshot collection \cite{RN382,RN418}, see the definition of $\alpha_i$ in Eq.\ \ref{eq11}. Figure \ref{fg12} shows that the modes selected by this criterion are prone to have a higher amplitude. The modes inside the unit circle are stable, and modes falling outside the unit circle are unstable. We note that some of the larger amplitude modes (meaning that their associated $\alpha_i$ values are high) are closer to the origin. This means that these modes have a strong decay rate, which means that their strong influences are only manifest at the initial state of the physical process. Hence, we can learn from Fig.\ \ref{fg12} that the AP method tends to select those fast-decaying modes which have a large amplitude. 
{\color{black}For the SP method the modes are selected based on the optimization constraint, see Eq.\ref{eq14}. In other words, the SP method determines which modes should be discarded so that the remaining ones can give the best performance.} Based this criterion, the SP method selects modes that reside close to the circle with low frequencies, keeping their influences in the whole physical evolution for a long time. This difference in selecting DMD modes between AP and SP method is more apparent as shown in Fig.\ \ref{fg12}(b) with 15 modes. The difference between the AP and SP is consistent with Jovanovic {\it et al.}\ \cite{RN382}. In the remainder of this section, we compare the different flow structures captured by the leading AP and SP modes.

\begin{figure}[!t]
	\centering
	\includegraphics[width=1\textwidth]{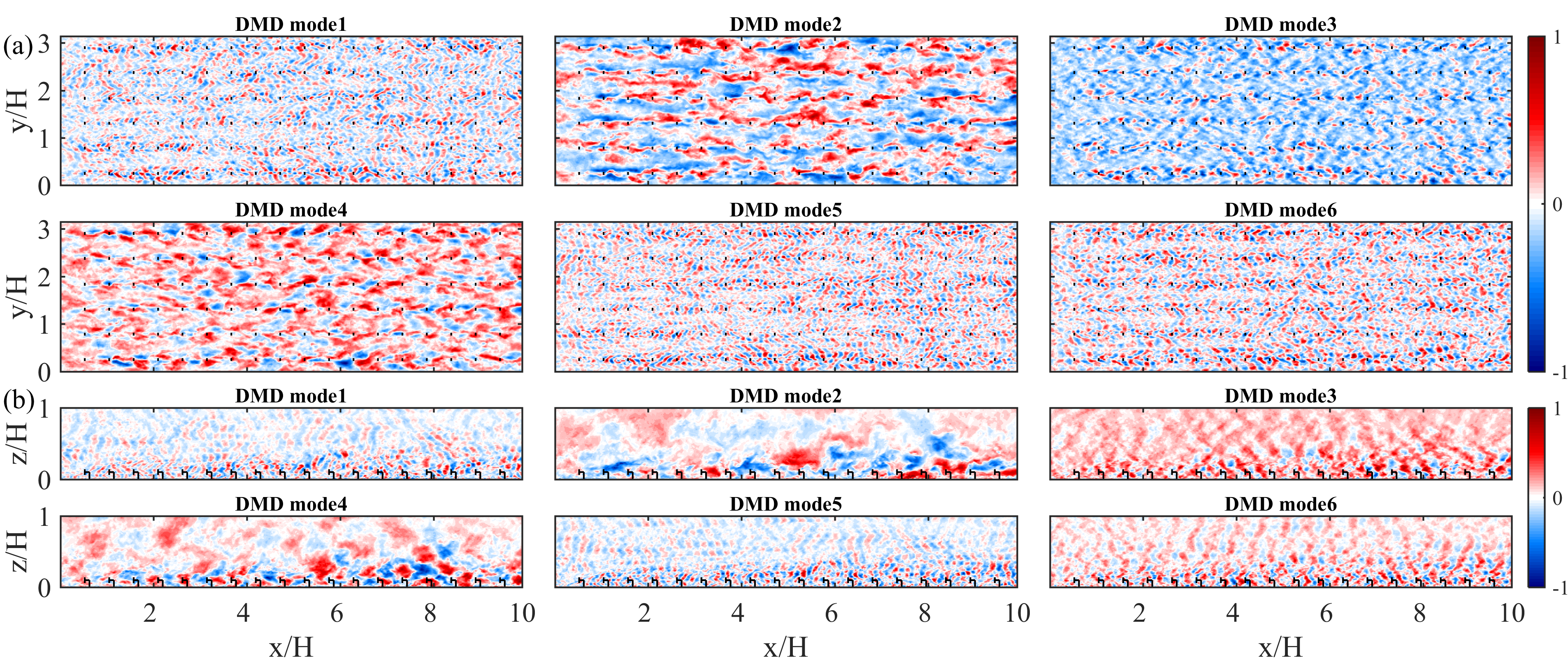}
	\includegraphics[width=1\textwidth]{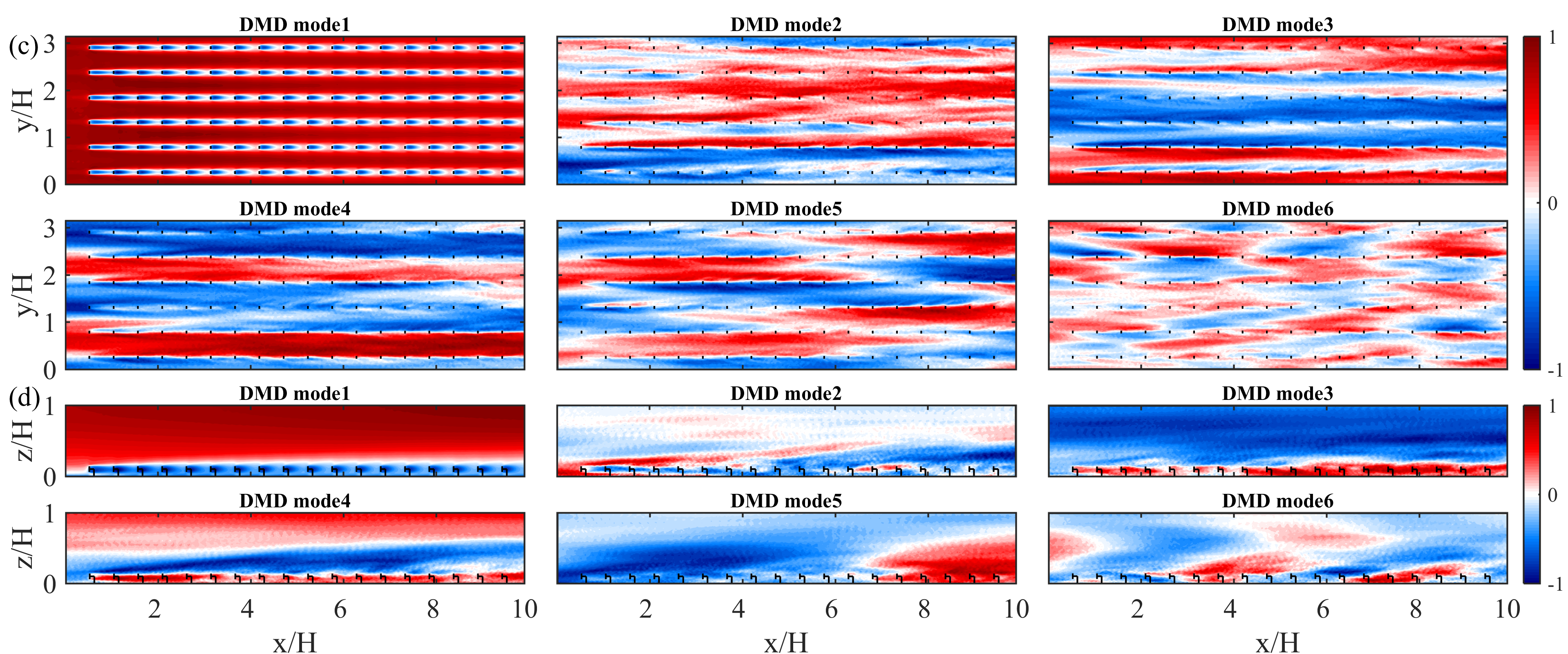}
	\caption{ DMD modes of `A' case (a)(b) by the AP method and (c)(d) SP method. XY-plane is located at $z/H=0.1$; XZ-plane is located at $y/D_t=1.31$}
	\label{fg13}
\end{figure} 
\begin{figure}[!ht]
	\centering
	\includegraphics[width=1\columnwidth]{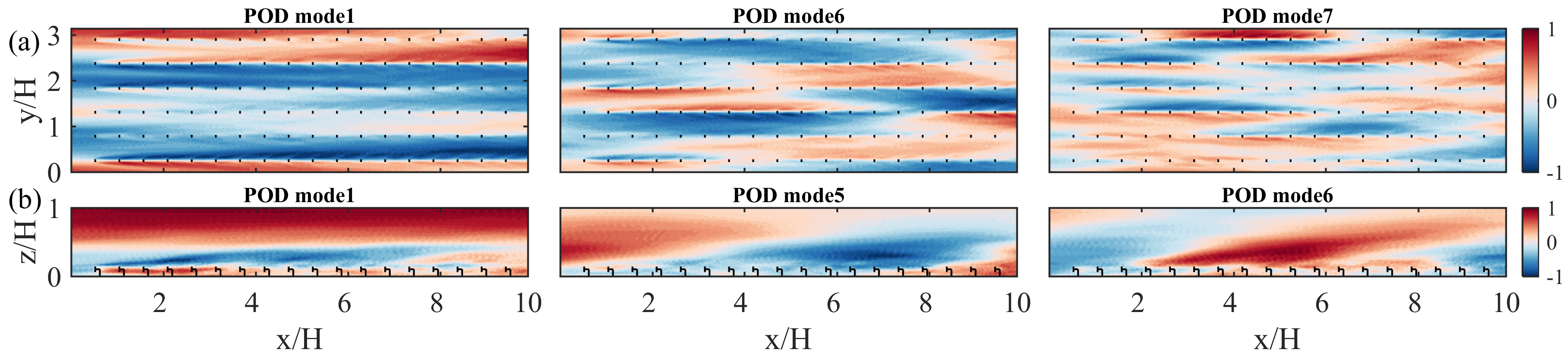}
	\caption{ POD modes of `A' case with top high energy (a) XY-plane at turbine hub-height $z/H=0.1$ (b) XZ-plane, $y/D_t=1.31$}
	\label{fig14}
\end{figure}

Fig.\ \ref{fg13}(a)(b) presents the leading DMD modes selected by the AP method and Fig.\ \ref{fg13}(c)(d) those obtained using the SP method. {\color{black} We note that, as mentioned above, the magnitude of the AP modes determines their ordering. The ordering of SP modes is determined by gradually raising $\gamma$ to see which modes remain. However, we find a set of 7 leading SP modes, which can not be further reduced even when high values of $\gamma$ (see Eq.\ref{eq14}) are used in the SP method. Therefore, here, and in the remainder of the paper, we decide to order these leading SP modes from low to high frequencies. The growth rate and frequencies of the SP medes presented in Fig.\ref{fig14} are given in Fig.\ref{fg12} the zoom-in panels.} The differences between the obtained modes are pronounced. Modes selected by the AP method focus on small-scale structures that decay rapidly. In contrast, modes chosen by the SP method tend to present large-scale flow structures. In particular, mode 2 obtained by the SP method captures the mean flow pattern discussed in section \ref{ssec2_1}, while the AP method completely ignores this mode. Besides, it is worth noting that the flow patterns captured by the SP method resemble the patterns grasped by POD modes of high energy, which are called streamwise roll structures \cite{RN313,zhang2020characterizing}. Some of these leading POD modes are presented in the panels of Fig.\ \ref{fig14} for reference. Just as the DMD modes, the POD modes are delimited by turbine rows in the spanwise direction and represent large-scale ABL structures. Besides, we note that POD mode 7 demonstrates the streamwise-varying rolls resulting from kinetic energy redistribution due to wake interaction. The POD result matches well with previous works \cite{RN313,zhang2020characterizing,RN444}.
From this point of view, we can understand the reason why DMD modes selected by the SP method are capable of reconstructing the original flow field in the wind farm more accurately than using the AP method. The reason is that the SP-DMD modes capture the long-term flow dynamics in the statistically steady state. That DMD modes resemble POD modes to some extent agrees with Tu {\it et al.}\ \cite{H_Tu_2014} who showed that the pattern of low-frequency DMD modes is similar to the leading POD modes.

\begin{figure}[!t]
	\centering
	\includegraphics[width=0.5\columnwidth]{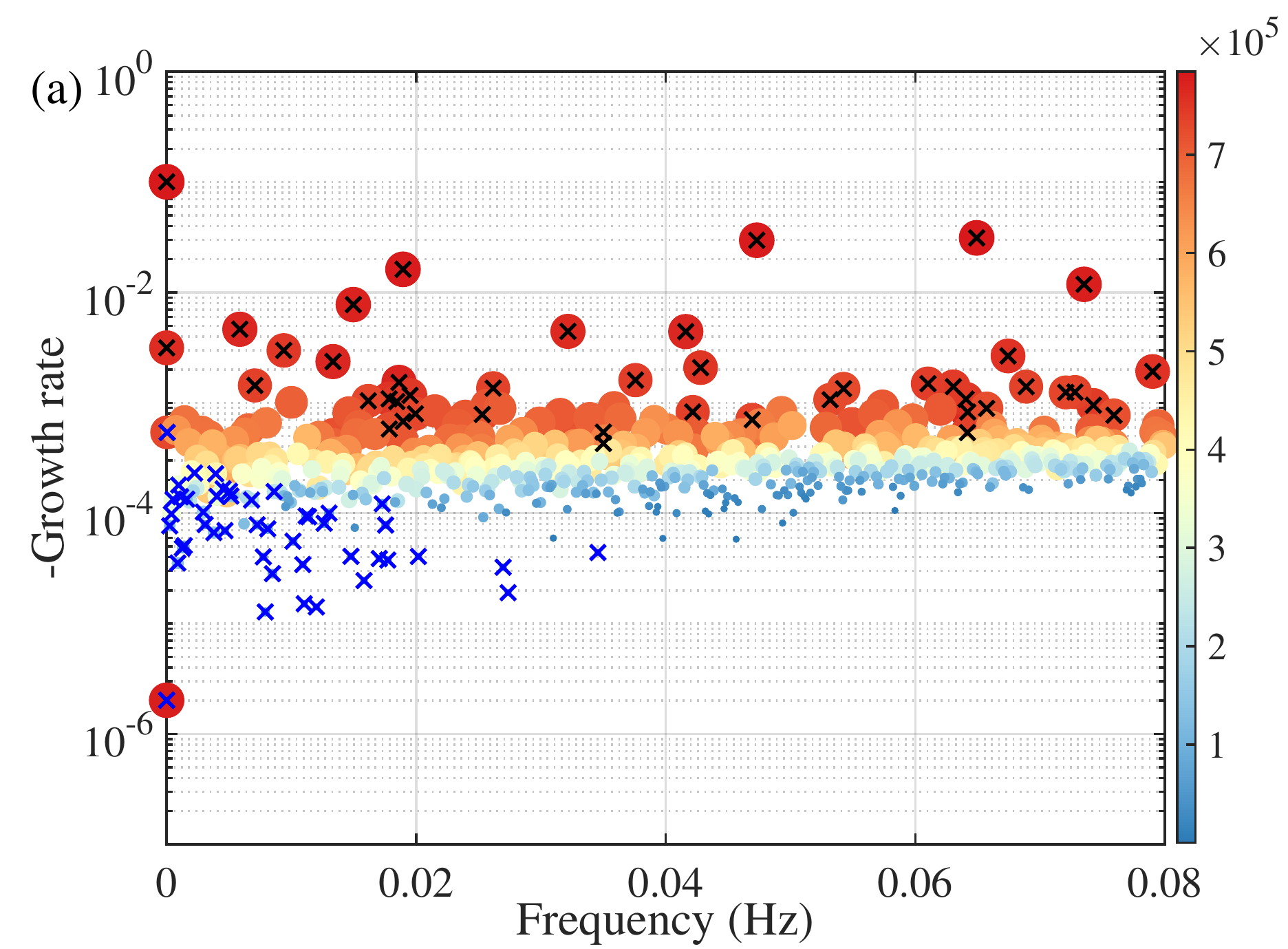}
	\caption{The spectrum of the DMD modes with $N_t=2500$. The amplitude of each DMD mode is represented by the size and the colour of the circle (see the colour bar). Circles superposed by black crosses: the first 200 modes selected by the AP method. Circles superposed by blue crosses: the first 200 modes selected by the SP method.}
	\label{fg15}
\end{figure}

Fig.\ \ref{fg15} presents the DMD modespectrum with the positive frequencies. Note that the frequency and growth/decay rate are calculated by Eq.\ \ref{eq18} below.
\begin{equation}
f_i=\frac{Im(\ln(\mu_i))}{2\pi \Delta t}, \ \ \ g_i=\frac{Re(\ln(\mu_i))}{\Delta t}
\label{eq18}
\end{equation}
where $\mu_i$ can found in Eq.\ \ref{eq10} and $Im$ and $Re$ indicate that the imaginary and real part of a complex number, respectively. The amplitude of each DMD mode is indicated by the circle size and the color, as shown in Fig.\ \ref{fg15}. In total, there are $1250$ modes ($N_t=2500$, but we only show half of the spectrum) with the size and the color of the circle (see the color bar) both presenting the amplitude of the DMD mode. The black crosses represent the first 200 modes selected by the AP method, and the blue crosses the first 200 modes selected by the SP method. Again, like the results in Fig.\ \ref{fg12}, one can clearly see the two methods favor different modes. The AP method tends to select the fast decaying modes with a high amplitude. In contrast, the SP method tends to select low-frequency modes with a low amplitude. Plus, most of the 200 modes chosen by the SP method sit within a low-frequency range below approximately $0.025\ Hz$. As for the AP method, we can see a wide distribution of scattered frequencies from 0 to around 0.08 $Hz$.
 Consequently, it can be concluded that the modes selected by the AP method are connected with the wake dynamics of a high frequency as supported by flow patterns in mode 1, mode 5, and mode 6 in Fig.\ \ref{fg13}(a)(b) where blue and red stripes are entangled indicating strongly oscillating small vortex structures shedding from the turbines. In contrast, the modes chosen by the SP method are mainly associated with low-frequency large-scale flow structures. For example, {mode 4 and mode 5 }in Fig.\ \ref{fg13}(c)(d) characterize the streamwise velocity deficit behind turbines, while mode 1 and mode 6 represent the large-scale structures similar to streamwise-varying rolls. Also, the frequency of mode 4 in Fig.\ \ref{fg13}(a)(b) is close to $f_t$ so we can see that its flow pattern is depicted by the alternating positive and negative perturbation whose characteristic length scale is close to $S_x$ (streamwise turbine spacing) in the streamwise direction is displayed in XZ-plane of Fig.\ \ref{fg13}(b). 

In this section, we compared the results obtained using the AP and SP methods for the aligned case `A'. 
{\color{black} The results for other layouts are attached in \ref{app1}. It is worthy stating that a comparison of the horizontally and vertically aligned case (Fig.\ref{fg13}) with the corresponding vertically staggered case (Fig.\ref{fg19}) shows that the DMD modes are different inside the wind farm. However, the structure of the DMD modes above the wind farm turns out to be quite similar for the wind farms with and without vertical staggering. The reason for this is that the large-scale flow structures above the wind farm formed in the atmospheric boundary layer are not considerably affected by the vertically staggering in the wind farm.}

\begin{figure}[!t]
	\centering
	\includegraphics[width=0.95\textwidth]{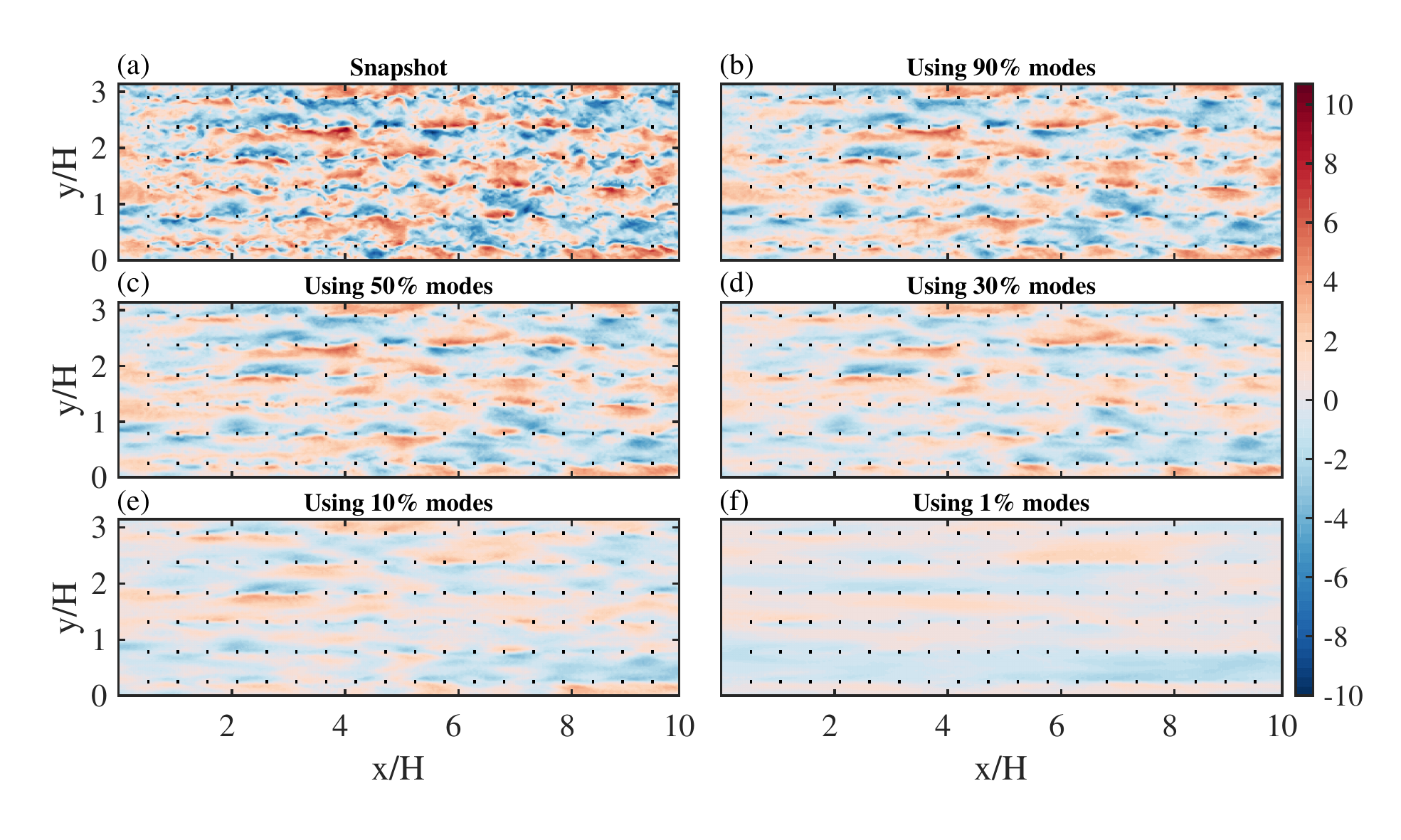}
	\caption{(a) Snapshot of fluctuation velocity from LES result; Fluctuation velocity field reconstructed from DMD modes chosen by the SP method excluding the mean mode : (b) 90\% modes; (c) 50\% modes; (d) 30\% modes; (e) 10\% modes; (f) 1\% modes. Note that as we use the same colorbar for all the figures, the amplitude of the fluctuation of the velocity naturally decreases when more DMD modes are discarded.}
	\label{fig2}
\end{figure}

\subsection{Flow reconstruction using DMD} \label{ssec3_33}

This section discusses the flow field reconstruction using the AP and SP methods based on Eq.\ \ref{eq11}. First, in order to gain a general understanding of how the DMD analysis performs, we reconstruct the fluctuating velocity component using a different number of the DMD mode by SP method, $1\%$, $10\%$, $20\%$, $30\%$, $50\%$ and $90\%$ of the available $N_t=2500$ DMD modes. As shown in Fig.\ \ref{fig2}, with the number of used modes decreasing, we can see a clear trend that when enough modes are used (e.g., 90\%), the flow reconstruction is faithful in terms of both flow structure and amplitude. It is encouraging to see that even 10\% of the modes can reconstruct the salient features of the large-scale flow structures in the fluctuation part of the velocity field (even though the amplitude of the fluctuation is smaller because the higher DMD modes are discarded). When using an even smaller number of DMD modes (1\% modes) as shown in panel (f), one can only get the streamwise rolls residing between the turbine rows, similar to what is obtained using POD \cite{RN313,zhang2020characterizing}. 

\begin{figure}[!t]
	\centering
	\includegraphics[width=0.9\textwidth]{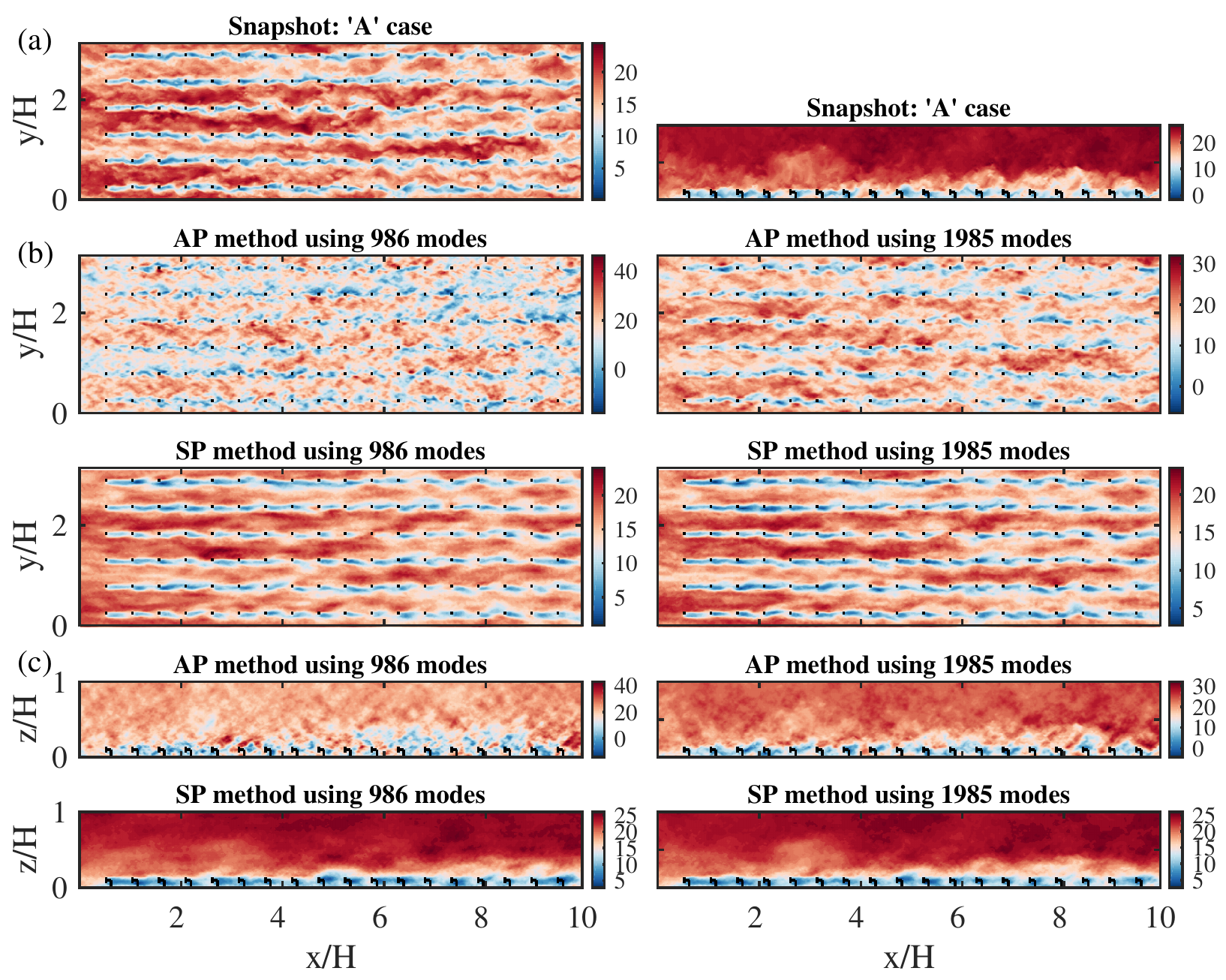}
	\caption{ (a) Snapshot from LES result; Reconstructed velocity field by AP and SP methods for `A' case: (b) XY-plane (c) XZ-plane. XY-plane locates at $z/H=0.1$; XZ-plane locates at $y/D_t=1.31$. }
	\label{fg17}
\end{figure}

Fig.\ \ref{fg17} shows the DMD reconstruction of the complete flow field (rather than the fluctuation part) for the AP and SP methods using 986 and 1985 modes out of $N_t=2500$. In the two cross-sections, we can see that the original snapshots manifest highly turbulent structures. The figure shows that the AP method apparently cannot reconstruct these flow structures with satisfactory accuracy, while the SP method captures the most salient flow structures correctly, even using a relatively limited number of modes. Obviously, some of the small-scale structures are inevitably lost by disregarding the higher DMD modes. \ref{app2} shows that similar finding are obtained for the other wind farm layouts.

\begin{figure}[!t]
	\centering
	\includegraphics[width=0.5\columnwidth]{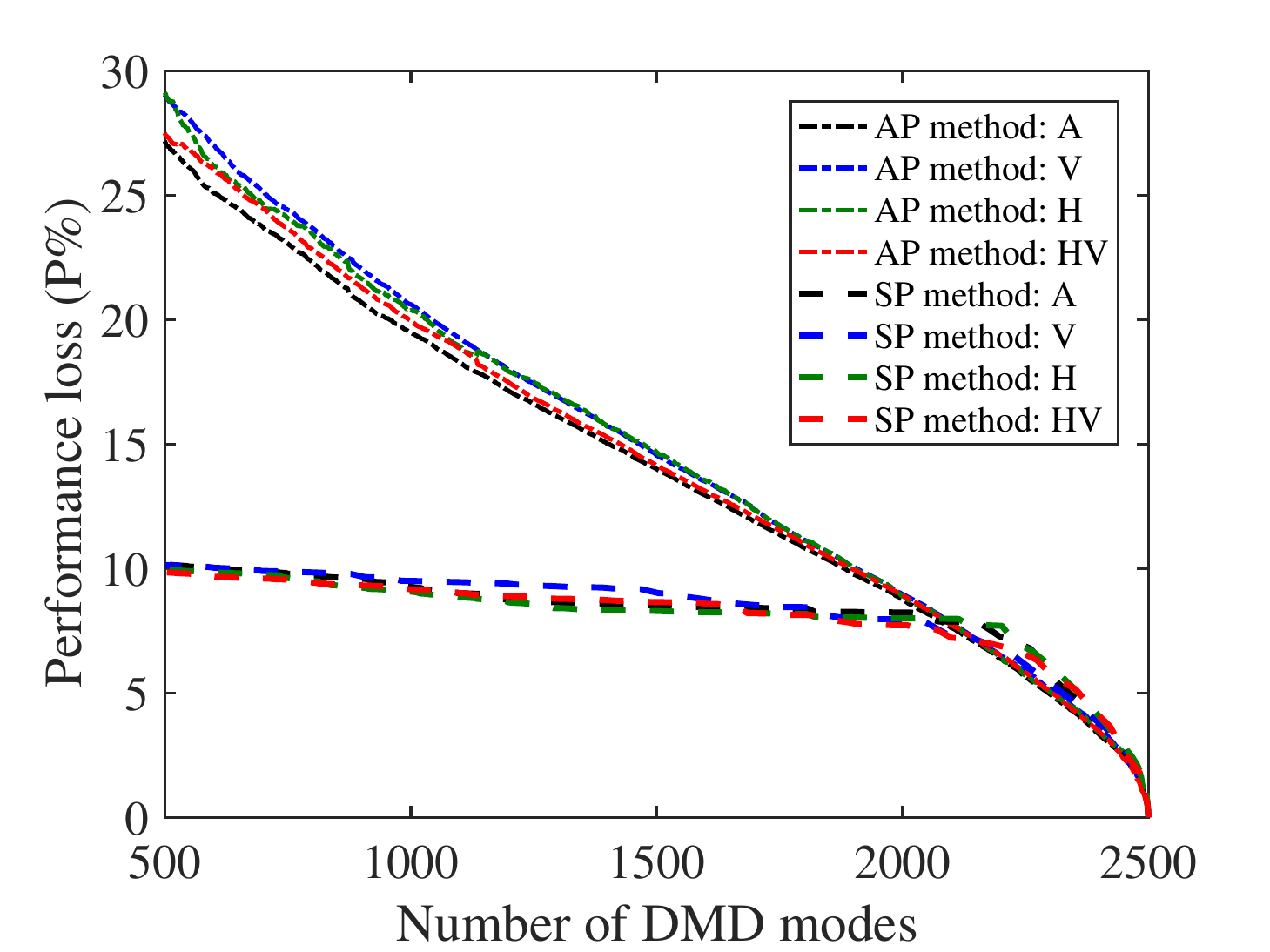}
	\caption{Performance loss as function of the number of DMD modes for the AP and SP methods.}
	\label{fg18}
\end{figure}

After gaining a visual impression of the performance of the AP and SP methods, we further quantify the performance of the methods by determining the performance loss, which is defined as \cite{RN382}
\begin{equation}
P\%=\frac{\left\|X-\Phi D_\alpha V_{and}\right\|_F}{\left\|X\right\|_F}\times100
\label{eq20}
\end{equation}
In Fig.\ \ref{fg18}, we show that increasing the number of DMD modes reduces the performance loss, i.e., the accuracy becomes higher for both the AP and SP method. When a limited number of DMD modes is retained, see the left end of the figure, the performance loss of the SP method is less than for the AP method. The performance loss of the AP method strongly depends on the number of retained modes. The reason is that the leading AP modes have a large decay rate, which does not capture the long-term system dynamics well. In contrast, the SP method prefers to select modes with a low decay rate, which captures the long-term dynamics better. In agreement with Fig.\ \ref{fg17} we find that retaining, e.g., 40\% or 80\% of the DMD modes in the SP method does not affect the performance loss much. Thus, the SP method can be considered superior to the AP method when a limited number of DMD modes is retained. We can also observe that the results for the four wind farm layouts are similar, even for the 'HV' case in which complicated flow patterns are formed. This indicates that the excellent performance of the SP method in the flow reconstruction is robust and general. 

\begin{figure}[h]
	\centering
	\includegraphics[width=0.5\textwidth]{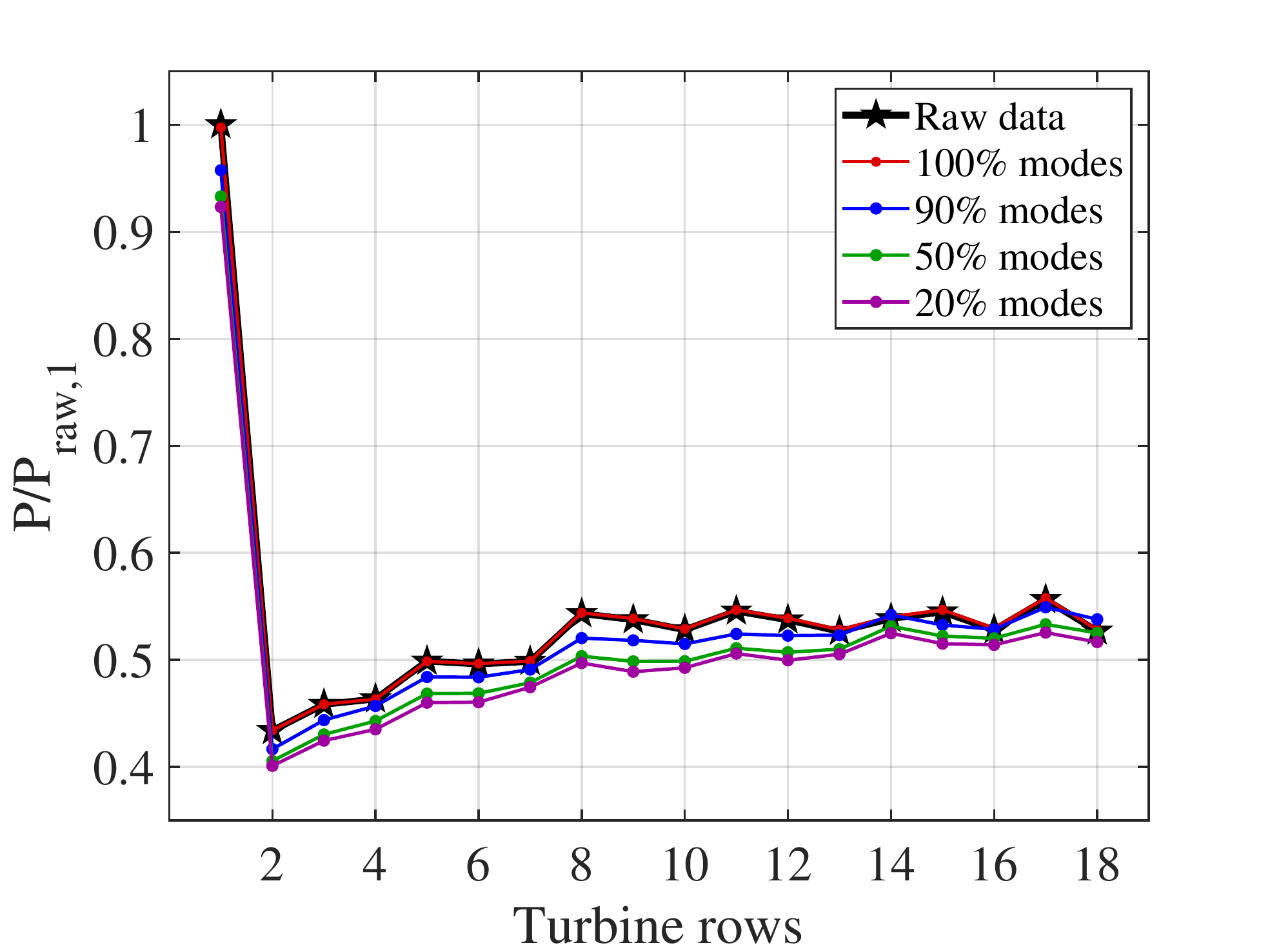}
	\caption{Power production normalized by the power of the first row in the original data, compared to that calculated using reconstructed flow field based on different number of DMD modes. Note that the mean mode is included in the reconstruction.}
	\label{Power_Reconstruction}
\end{figure}

{\color{black} Finally, we attempt to use different numbers of DMD modes to compute the power production for the `A' case by the SP method. First, the flow field is reconstructed using different numbers of DMD modes ($100\%, 90\%, 50\%$ and $20\%$) with the mean mode included. Then, $U_d$, the disk-average velocity, is obtained by conducting disk-averaging of the reconstructed flow field for different numbers of DMD modes, based on which the power production is calculated correspondingly. Fig.\ \ref{Power_Reconstruction} shows the normalised power production by the first row using the raw data, compared to the power production calculated using the DMD-reconstructed flow field. It demonstrates that a limited number of DMD modes allow for an accurate reproduction of the power production data once the mean mode is included in the reconstruction.}


\section{Conclusions}\label{sec4}

In this work, LES is used to study the turbulent flow structures in various wind farm layouts using DMD. The combination of horizontal and vertical staggering (the `HV' case) has not been considered before, and our results indicate that it generates a higher power production than either vertical or horizontal staggering case. Subsequently, the turbulent flow structures in the wind farm are then analyzed by DMD. It is noted that unlike previous DMD studies, which focused on wake dynamics behind a single turbine or a small number of turbines \cite{RN425,RN420,RN374}, we concentrate our analysis on the complex flow dynamics in extended wind farms. We used two methods to select the DMD modes and found that the AP method tends to select modes with a high frequency. The frequencies of the leading modes can be as high as $0.08\ Hz$ and these modes characterize the highly oscillating flow patterns corresponding to the small-scale vortex structures shedding from turbines. On the other hand, the SP method tends to select modes with dominant large-scale flow structures with low frequencies and decay rates. The flow patterns of the modes selected by the SP method are {\it similar} to the dominant flow structures identified by POD analysis, see also Ref. \cite{RN313,zhang2020characterizing}. However, it is essential to emphasize that DMD, in contrast to POD, can resolve the temporal evolution of the structures. 

The temporal information in DMD can be used to reconstruct the time evolution of the flows. To assess the potential of DMD for the development of reduced-order models that are suitable to study wind farm dynamics and study wind farm control strategies, we investigated how well a subset of DMD modes selected by either the AP or SP method can reconstruct the flow field in a wind farm. We find that the SP method provides higher accuracy in global physical space while the AP method offers more information on small-scale structures. It is promising that less than $10\%$ of the SP modes allow for an adequate representation of the flow field in the wind farm. This highlights the great potential of DMD for flow analysis, prediction, and wind farm control. As the flows, we simulated here are highly turbulent, it is reasonable to observe that the nonlinearity makes it difficult to accurately capture the fluctuating velocity field using a small number of DMD modes. It also has to be mentioned that the DMD reconstruction works almost equally well for all the turbine layouts considered, showcasing the robustness of DMD in dealing with the complex turbulent flow structures created in wind farms.

To conclude, our work fills the gap of flow analysis of complex wake dynamics in large numerically simulated wind farms using DMD. There is an enormous potential and need for studies on DMD-based flow control for improving wind farm efficiency or more accurate flow field reconstruction.

\section*{Acknowledgement}
We acknowledge doctoral research scholarships from National University of Singapore and the Tier-1 grant from the Ministry of Education, Singapore (MOE WBS no.R-265-000-654-114). This work is also part of the Shell-NWO/FOM-initiative Computational sciences for energy research of Shell and Chemical Sciences, Earth and Live Sciences, Physical Sciences, FOM, and STW and an STW VIDI grant (No. 14868).

\clearpage

\appendix

\section{DMD modes for H,V, and HV cases}\label{app1}

This appendix presents the DMD modes for the wind farms with other turbine layouts (horizontal staggering, vertical staggering, and horizontal-vertical staggering). In the following analyses, we will briefly summarize the similarities and differences between different cases.

\subsection{DMD modes for `V' case}\label{app1_V}
\begin{figure}[!ht]
	\centering
	\includegraphics[width=1\textwidth]{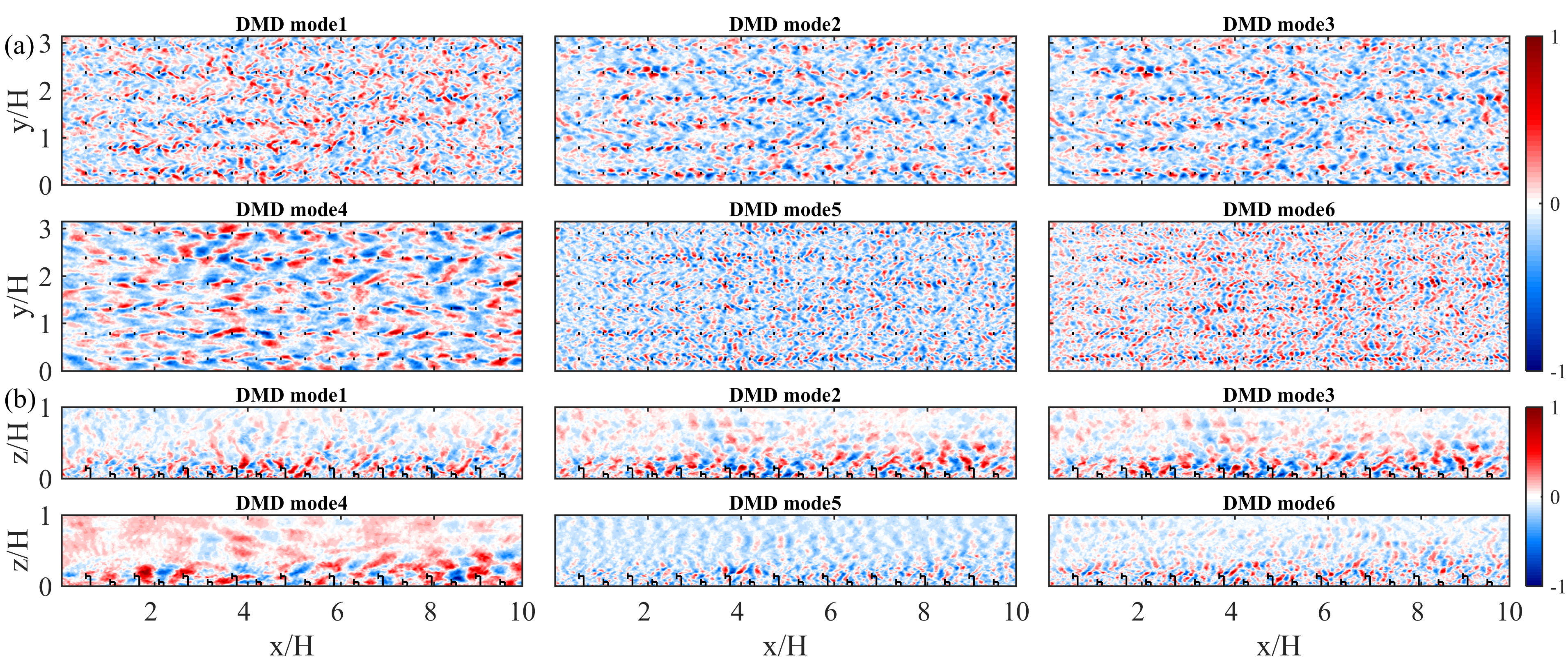}
	\includegraphics[width=1\textwidth]{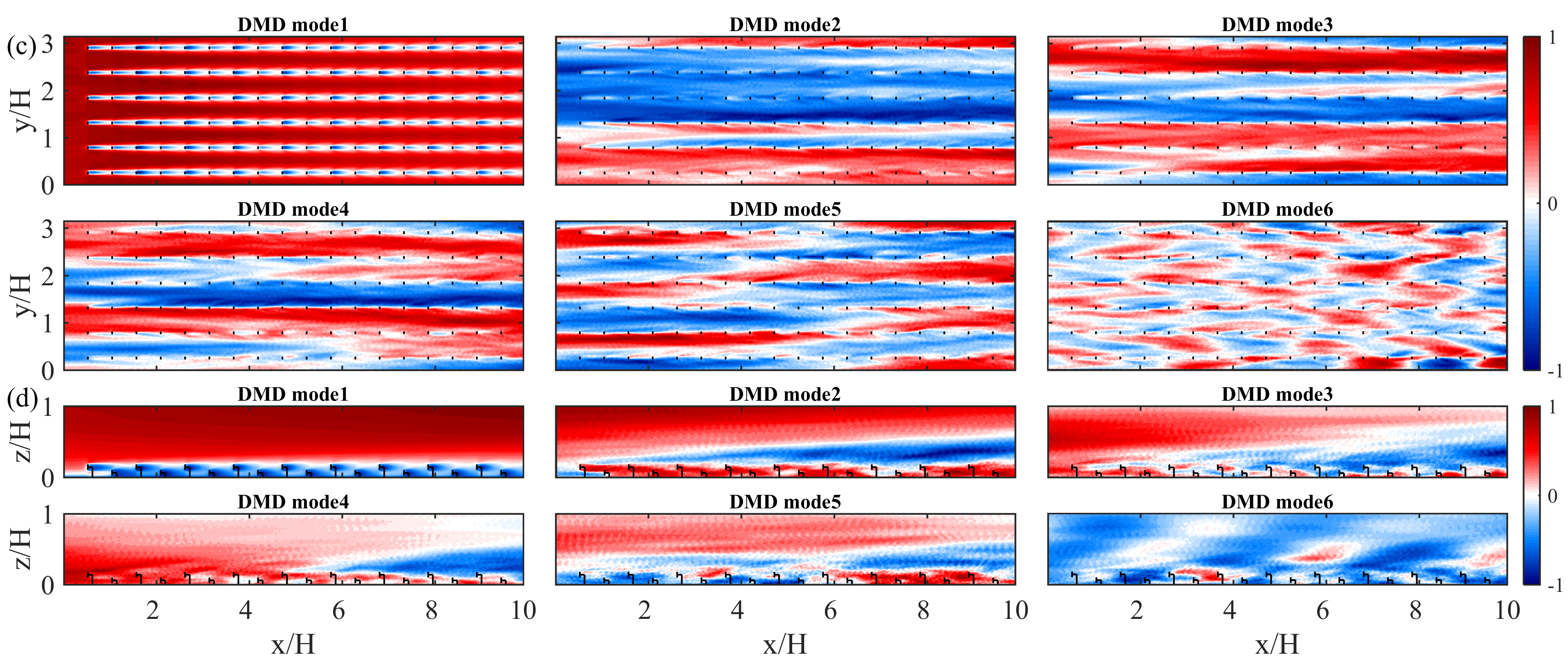}
	\caption{Selected DMD modes for the `V' case (a)(b) by the AP method and (c)(d) SP method. XY-plane is located at $z/H=0.1$; XZ-plane is located at $y/D_t=1.31$.}
	\label{fg19}
\end{figure}

\begin{figure}[!ht]
	\centering
	\includegraphics[width=1\textwidth]{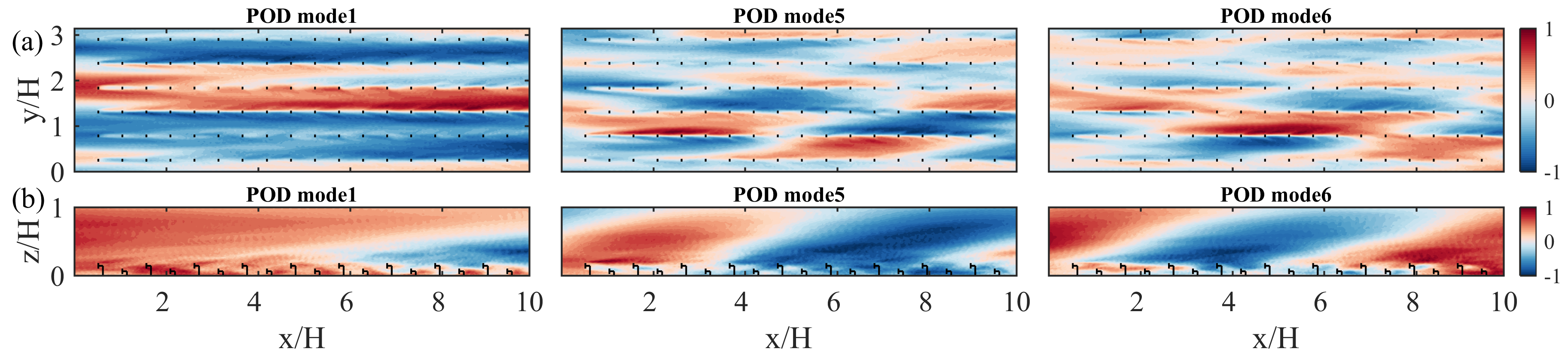}
	\caption{ Selected leading POD modes for the `V' case. (a) XY-plane at turbine height, $z/H=0.1$. (b) XZ-plane with $y/D_t=1.31$.}
	\label{fg20}
\end{figure}

Fig.\ \ref{fg19} shows the DMD modes for the `V' case using the AP (panels a,b) and SP methods (panels c,d). The features of the `V' case are similar to that of the `A' case. The flow pattern of mode 4 in Fig.\ \ref{fg19}(a)(b) shows that the characteristic length scale corresponds to the turbine spacing $S_{x}$. The frequency of this mode nearly equals $f_t$. The flow pattern alternates between positive and negative perturbations roughly on the turbine disk. In Fig.\ \ref{fg19}(c)(d), the DMD modes selected by the SP method characterize large-scale structures. The velocity deficit behind turbines is clearly illustrated in Fig.\ \ref{fg19}(d) mode 3-5. In mode 4 of Fig.\ \ref{fg19}(d), the velocity deficit pattern varies with turbine height in vertical direction induced by vertical staggering. DMD mode 1 in Fig.\ \ref{fg19}(c)(d) characterizes the flow pattern of a fluid parcel passing multiple turbine spacing, which is similar to the streamwise-varying roll structures. The flow patterns captured using the SP method also resemble those in POD modes shown in Fig.\ \ref{fg20}.

\subsection{DMD modes for `H' case}
\begin{figure}[!ht]
	\centering
	\includegraphics[width=1\textwidth]{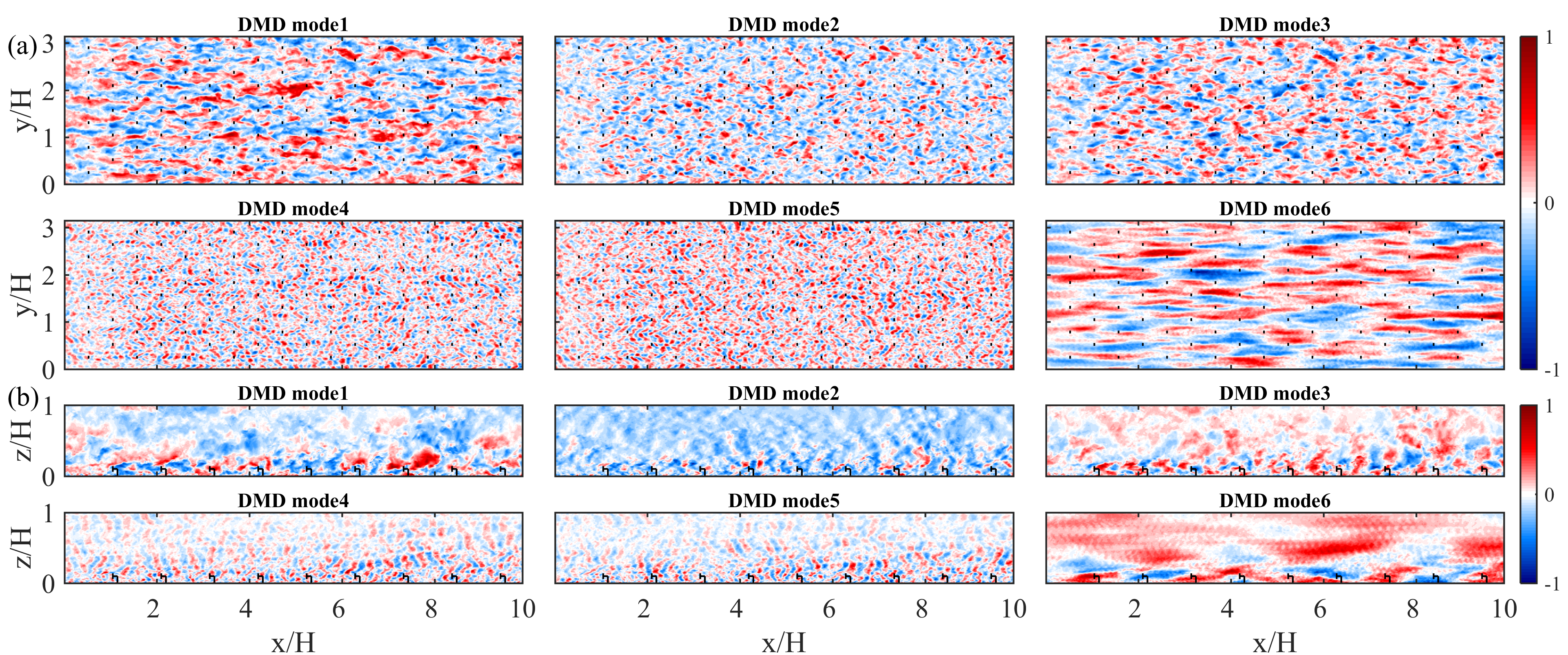}
	\includegraphics[width=1\textwidth]{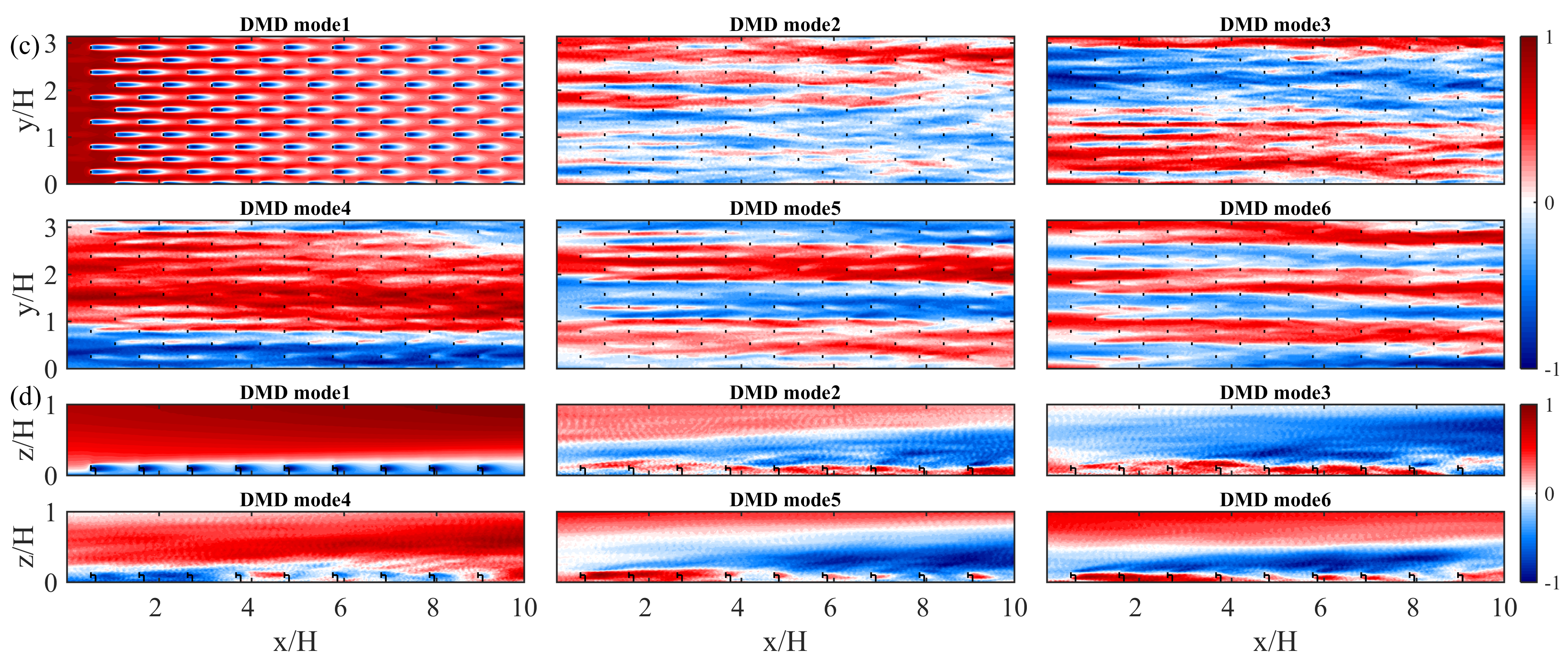}
	\caption{ Selected DMD modes for the `H' case (a)(b) by the AP method and (c)(d) SP method. XY-plane is located at $z/H=0.1$; XZ-plane is located at $y/D_t=1.57$.}
	\label{fg21}
\end{figure}

\begin{figure}[!ht]
	\centering
	\includegraphics[width=1\textwidth]{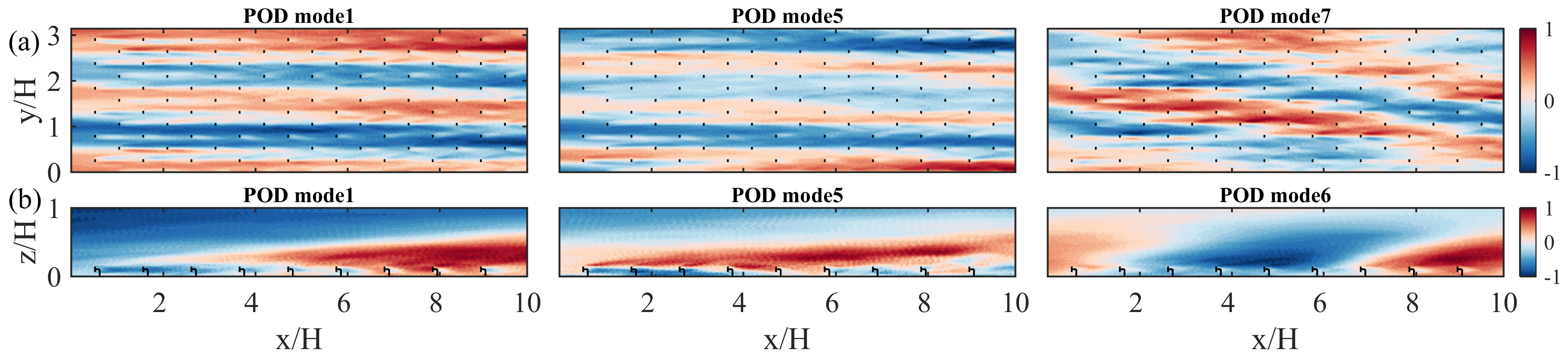}
	\caption{ Selected leading POD modes for the `H' case. (a) XY-plane at turbine height, $z/H=0.1$ (b) XZ-plane with $y/D_t=1.57$.}
	\label{fg22}
\end{figure}

In the `H' case, one can see that mode 6 of Fig.\ \ref{fg21}(a)(b) characterizes the length scale for a fluid parcel traveling through the turbine spacing. However, due to horizontal staggering, the frequency of mode 6 decreases to roughly $0.5f_t$. So, the flow pattern of mode 6 in Fig.\ \ref{fg21}(a)(b) reveals prolonged stripes spanning two streamwise turbines as compared with mode 4 of Fig.\ \ref{fg13}(a)(b) in `A' case and mode 4 of Fig.\ \ref{fg19}(a)(b) in `V' case. The result is consistent with the turbine layouts. The remaining DMD modes shown in panels (a,b) features even smaller length scales. Large-scale flow structures are captured by the SP method as illustrated in Fig.\ \ref{fg21}(c)(d), which is related to a fluid parcel traveling through streamwise distance longer than the turbine spacing (longer length scales). The patterns of DMD modes chosen by the SP method also resemble those of POD modes of this case in Fig.\ \ref{fg22}. What is different is the roll structures in POD modes are re-distributed and no longer delimited by turbine rows, for the turbines are horizontally staggered.

\subsection{DMD modes for `HV' case}
\begin{figure}[!t]
	\centering
	\includegraphics[width=1\textwidth]{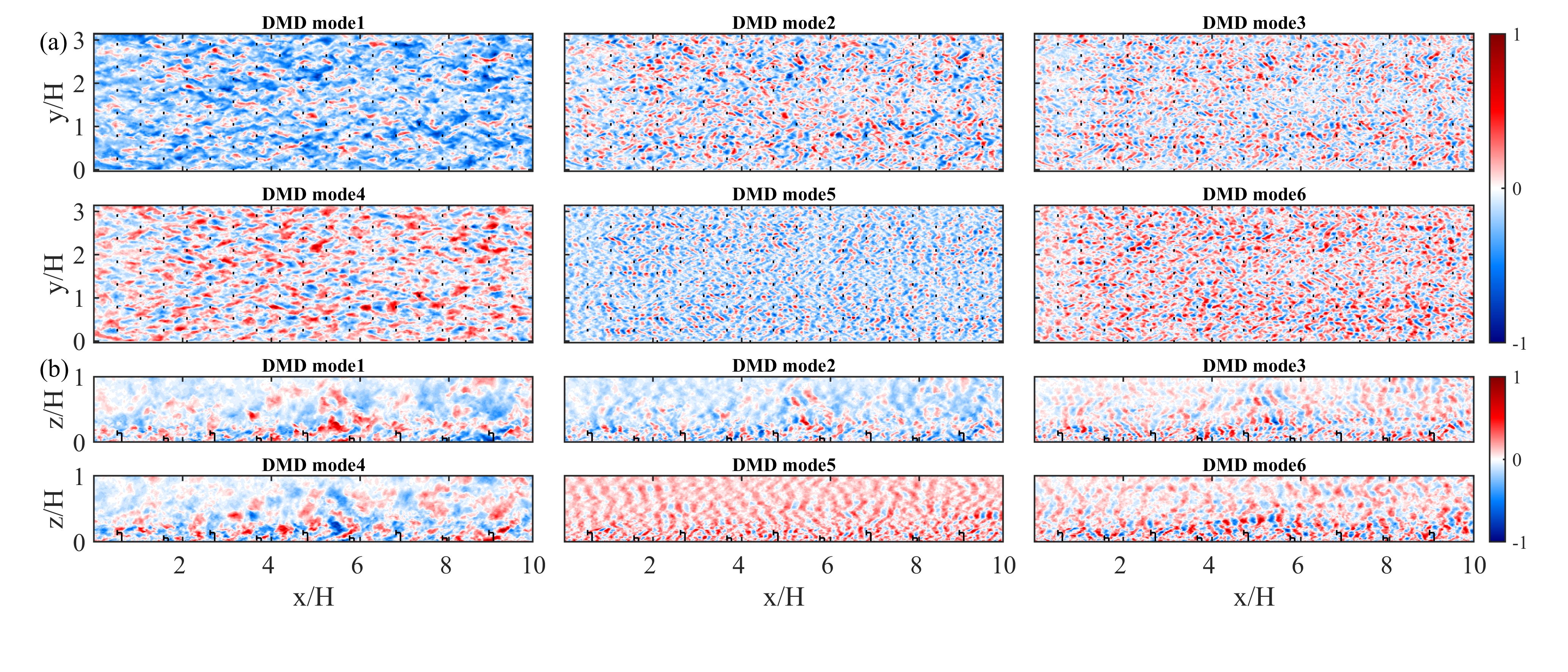}
	\includegraphics[width=1\textwidth]{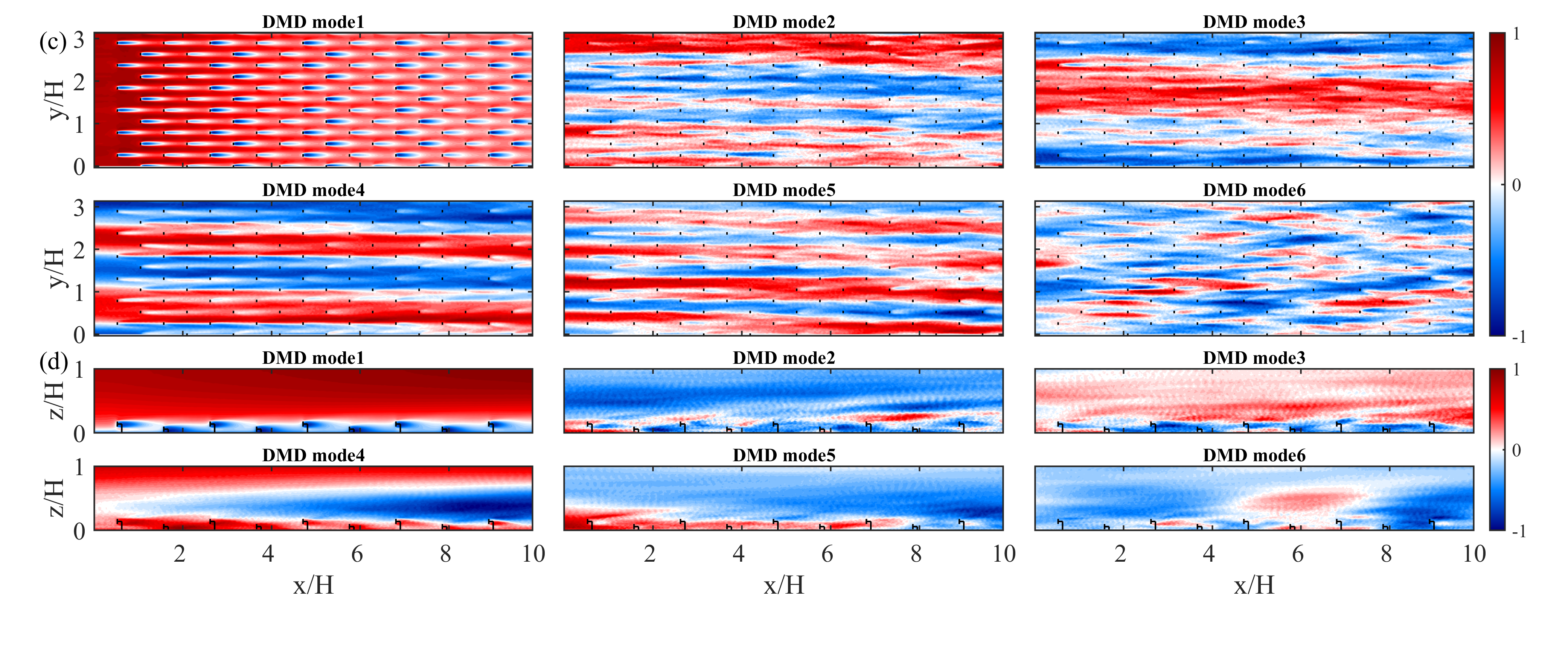}
	\caption{ Selected DMD modes for the `HV' case (a)(b) by the AP method and (c)(d) SP method. XY-plane is located at $z/H=0.1$; XZ-plane is located at $y/D_t=1.57$.}
	\label{fg23}
\end{figure}

\begin{figure}[!t]
	\centering
	\includegraphics[width=1\textwidth]{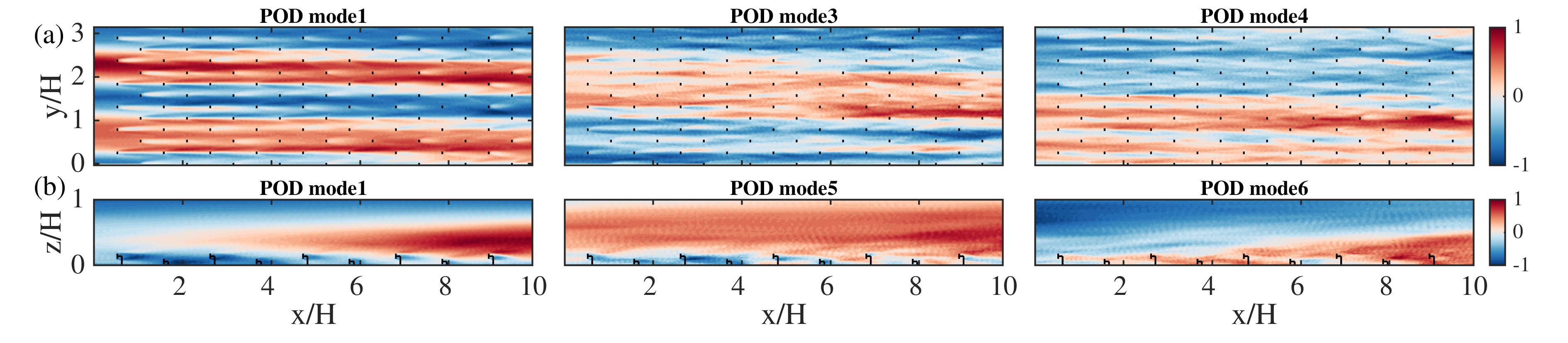}
	\caption{Selected leading POD modes for the `HV' case. (a) XY-plane at turbine hub-height, $z/H=0.1$ (b) XZ-plane with $y/D_t=1.57$.}
	\label{fg24}
\end{figure}

Similarly, in the `HV' case, the AP method tends to pick out high-amplitude modes while the SP method leads to selecting more persistently influential modes. The DMD modes obtained using the AP method show strongly oscillating flow patterns while the flow patterns determined by the SP method conform with the large-scale flow structures. The flow pattern whose characteristic length scale approximately equals 2 $S_{x}$, is also discovered in Fig.\ \ref{fg23}(a)(b) mode 6. The frequency of this mode is adjacent to that of mode 6 in the `H' case, which is roughly $0.5f_t$. Fig.\ \ref{fg23}(c)(d) reveal the large-scale structures captured by SP method. The velocity deficit behind turbines is shown in mode 4 and mode 5. Streamwise varying-roll structures are captured in mode 3 and mode 6. These flow patterns resemble those of POD shown in Fig.\ \ref{fg24} to some extent.

\clearpage




\clearpage

\section{Flow field reconstruction}\label{app2}
\begin{figure}[!ht]
	\centering
	\includegraphics[width=0.65\textwidth]{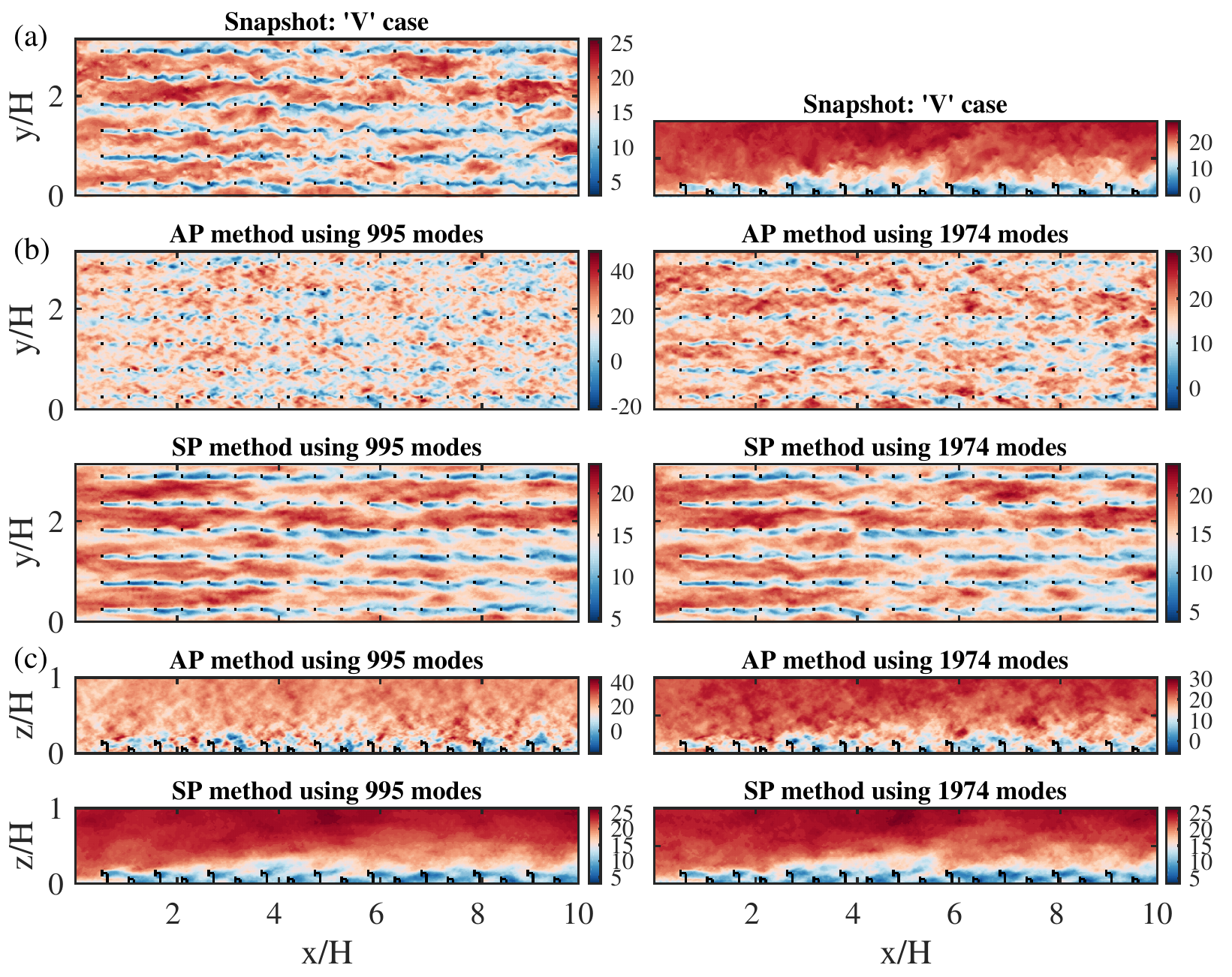}
	\caption{ (a) Snapshot from LES. Reconstructed velocity field using the AP and SP methods for the ‘V’ case: (b) XY-plane with $z/H=0.1$; (c) XZ-plane with $y/D_t=1.31$.}
	\label{fg28}
\end{figure}

\begin{figure}[!ht]
	\centering
	\includegraphics[width=0.65\textwidth]{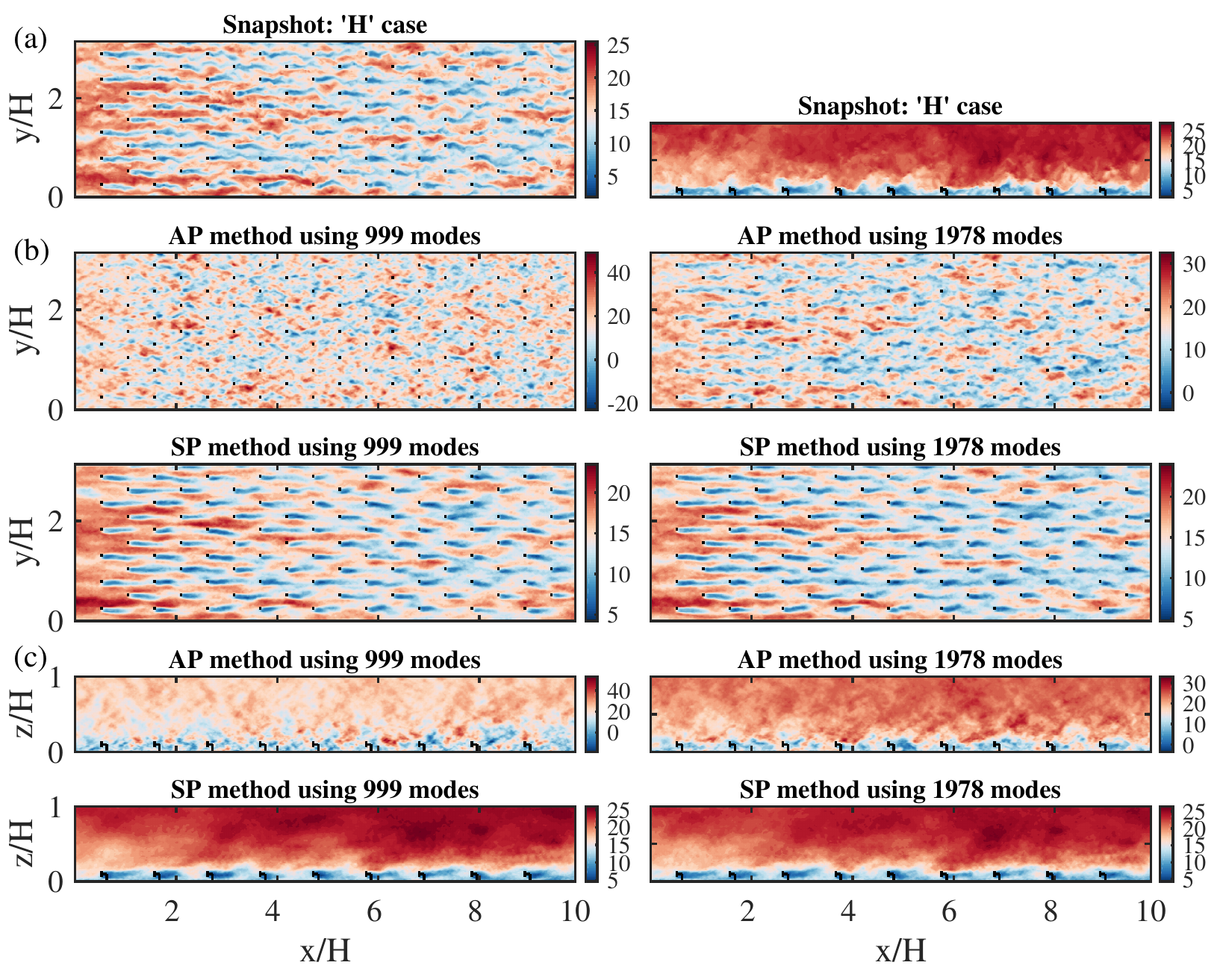}
	\caption{ (a) Snapshot from LES. Reconstructed velocity field using the AP and SP methods for the ‘H’ case: (b) XY-plane with $z/H=0.1$; (c) XZ-plane with $y/D_t=1.57$.}
	\label{fg29}
\end{figure}

\begin{figure}[!ht]
	\centering
	\includegraphics[width=0.65\textwidth]{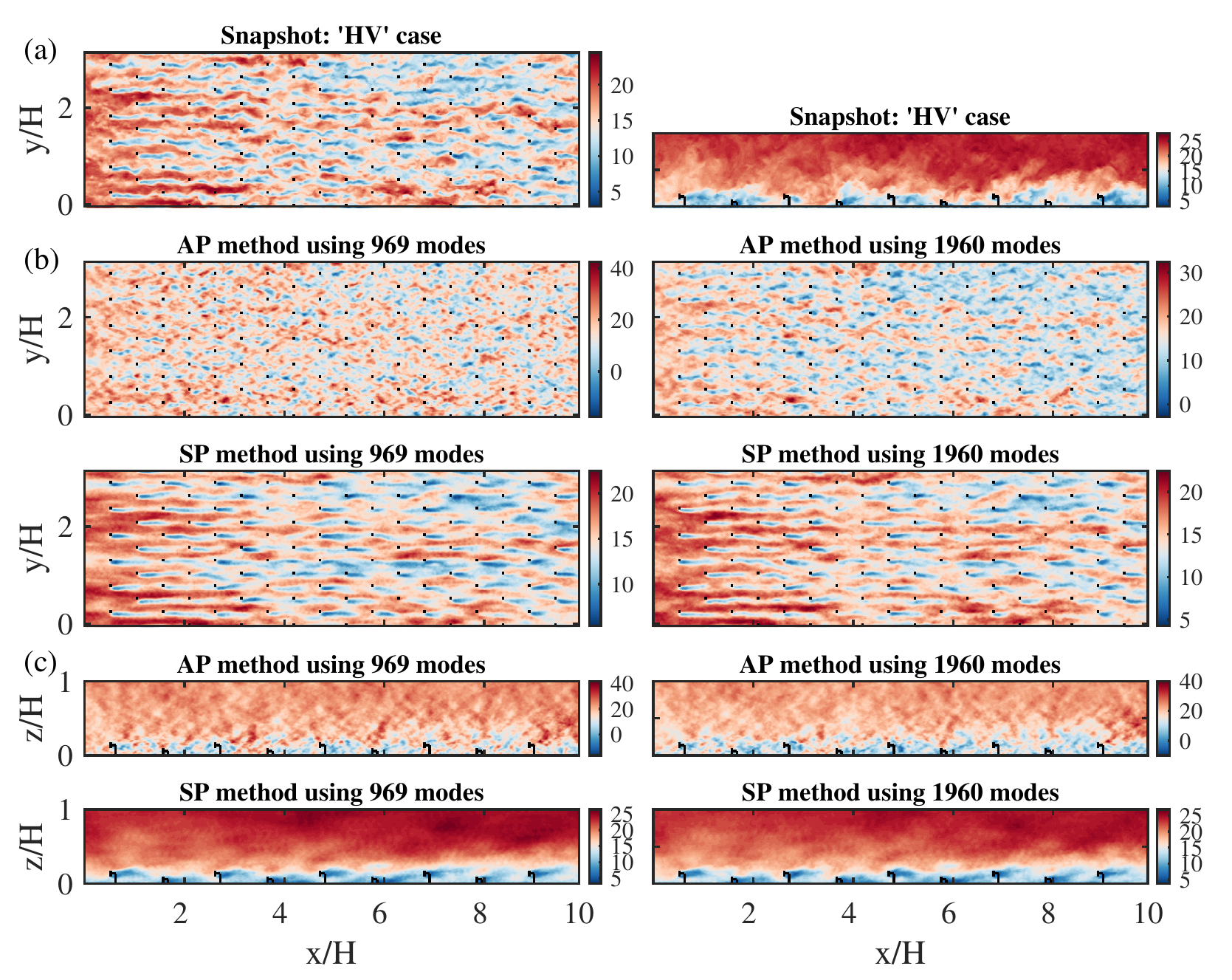}
	\caption{ (a) Snapshot from LES. Reconstructed velocity field using the AP and SP methods for the ‘HV’ case: (b) XY-plane with $z/H=0.1$; (c) XZ-plane with $y/D_t=1.57$.}
	\label{fg30}
\end{figure}

The flow field reconstruction results of the cases 'V', 'H', and 'HV' are presented below using the AP and SP method (roughly 40\% and 80\% of the total $N_t=2500$ modes). The results from these pictures are consistent with the conclusion in the main text: the AP method tends to capture the original flow field's small-scale structures while the SP method enables one to acquire the global flow field accurately. This shows that the SP-based DMD is capable of reconstructing the different original flow features for different layouts. The velocity deficit of the horizontally staggered layouts is more evident in the reconstructed snapshot as shown in Fig.\ \ref{fg29} and Fig.\ \ref{fg30}. Conversely, the wake effects are more evident in Fig.\ \ref{fg28}.


\clearpage

\bibliographystyle{ieee} 
\bibliography{windfarm_rs}

\end{document}